\documentclass[preprint,pre,floats,showkeys,aps,amsmath,amssymb,superscriptaddress,12pt,nofootinbib]{revtex4-1}
\usepackage{graphicx}
\usepackage{natbib}
\setcitestyle{square,comma,numbers,sort&compress}
\usepackage{amsmath,amsfonts,amssymb}
\usepackage{array}
\usepackage{tabularx}
\usepackage{enumerate}
\usepackage[left = 1.0in, top=1.0in,right=1.0in,bottom=1.0in,a4paper]{geometry}
\usepackage[dvipsnames]{xcolor}
\usepackage{colortbl}
\usepackage{comment}
\usepackage{siunitx}
\usepackage{lineno}
\usepackage[caption=false]{subfig}
\usepackage{tikz}
\usepackage[compat=1.1.0]{tikz-feynman}
\usepackage{cancel}
\usepackage{hyperref}

\newcommand{\be}{\begin{equation}}
\newcommand{\ee}{\end{equation}}
\newcommand{\ben}{\begin{equation*}}
\newcommand{\een}{\end{equation*}}

\def\bea{\begin{eqnarray}}
\def\eea{\end{eqnarray}}
\def\bean{\begin{eqnarray*}}
\def\eean{\end{eqnarray*}}

\def\l2{\log_2\,}
\newcommand{\barr}{\begin{array}}
\newcommand{\earr}{\end{array}}

\newcommand{\bed}{\begin{displaymath}}
\newcommand{\eed}{\end{displaymath}}
\newcommand{\bal}{\begin{array}{ll}}
\newcommand{\eal}{\end{array}}

\def\sc#1{\textstyle{\scriptstyle #1}}

\def\mc#1{\mathcal#1}

\newcolumntype{Y}{>{\centering\arraybackslash}X}
 
\newcommand{\g}[1]{\mathbf{#1}}
\newcommand{\vev}[1]{\langle #1 \rangle_0}
\newcommand{\gb}[1]{\bar{\mathbf{#1}}}

\newcommand{\blue}[1]{\textcolor{black}{#1}}

\newcommand{\myboxed}[1]{%
  \rlap{\hspace*{\dimexpr\fboxrule+\fboxsep\relax}%
    \phantom{\m@th$\displaystyle#1$}}%
    \smash{\boxed{#1}}}

\renewcommand{\thesection}{\arabic{section}{\Large}}
\renewcommand{\thesubsection}{\thesection.\arabic{subsection}}
\renewcommand{\thesubsubsection}{\thesubsection.\arabic{subsubsection}}

\makeatletter
\renewcommand{\p@subsection}{}
\renewcommand{\p@subsubsection}{}
\makeatother

\begin{document}

\title{\Large Leptogenesis from the Asymmetric Texture \\\vspace*{1cm}}

\author{Moinul Hossain Rahat} 
\email[Email: ]{mrahat@ufl.edu}
\affiliation{\small{Institute for Fundamental Theory, Department of Physics,
University of Florida, Gainesville, FL 32611, USA }\\\vspace*{1cm}}



\begin{abstract}
\vskip 0.5cm
We investigate non-resonant thermal leptogenesis in the context of the $SU(5) \times \mc T_{13}$ ``asymmetric texture'', where both Dirac and Majorana ${CP}$ violation arise from a single phase in the tribimaximal seesaw mixing matrix. We show that the baryon asymmetry of the universe can be explained in this model only when flavor effects are considered for right-handed neutrino masses of $\mc O(10^{11} - 10^{12})$ $\text{GeV}$. The sign of the baryon asymmetry also determines the sign of the previously predicted Dirac $\cancel{CP}$ phase $|\delta_{CP}| = 1.32\pi$, consistent with the latest global fit $\delta_{CP}^{PDG} = 1.37 \pm 0.17\pi$.
\end{abstract}

\maketitle


\section{Introduction}
The observable lepton mixing in the Pontecorvo-Maki-Nakagawa-Sakata (PMNS) matrix has been measured to contain two large and one small angle, unlike the nearly-identity 
Cabibbo-Kobayashi-Maskawa (CKM) matrix \cite{pdglive, deSalas:2020pgw, *Esteban:2020cvm}. This inspires contemplating the large mixing angles originating from the unknown $\Delta I_w = 0$ physics, whereas the small reactor angle comes entirely from the ``Cabibbo haze'' \cite{Datta:2005ci, *everett2006viewing, *Everett:2006fq, *Kile:2013gla, *Kile:2014kya} of the $\Delta I_w = \frac{1}{2}$ sector. This idea is implemented in the $SU(5)$ ``asymmetric texture'' \cite{Rahat:2018sgs} where the $\Delta I_w = 0$ seesaw matrix is assumed to be diagonalized by the tribimaximal (TBM) mixing \cite{harrison2002tri, *tbm2, *xing2002nearly, *he2003some, *wolfenstein1978oscillations, *Luhn:2007sy} with a $\cancel{CP}$ phase. \blue{The asymmetry, introduced minimally in the down-quark and the charged-lepton Yukawa matrices, is essential to explain the reactor angle and it determines the TBM phase up to a sign}. This single phase brings all three lepton mixing angles within $3\sigma$ of their Particle Data Group (PDG) value and predicts ${CP}$ violation in the lepton sector consistent with the current global fits \cite{pdglive, deSalas:2020pgw, *Esteban:2020cvm}. 

The asymmetry of the texture singles out $\mc T_{13} \equiv \mc Z_{13} \rtimes \mc Z_3$ \cite{bovier1981finite, *bovier1981representations, *fairbairn1982some, *ding2011tri, *hartmann2011neutrino, *Hartmann:2011dn, *ishimori2012introduction, *Ramond:2020dgm}, an order 39  discrete subgroup of $SU(3)$, as the smallest family symmetry. The electroweak sector of the texture is explained in an $SU(5) \times \mathcal{T}_{13}$ model in Ref.~\cite{Perez:2019aqq} and its seesaw sector is explored in Ref.~\cite{Perez:2020nqq}. Guided by minimality in the particle content and simplicity in the vacuum structure of the scalars, this model yields the normal ordering of light neutrino masses such that $m_{\nu_1} = 27.6$, $m_{\nu_2} = 28.9$ and $m_{\nu_3} = 57.8$ $\text{meV}$ through the seesaw mechanism involving four right-handed Majorana neutrinos. \blue{The sum of these masses almost saturates the Planck bound $\sum_i m_i \leq 120\ \text{meV}$ \cite{aghanim2018planck, Vagnozzi:2017ovm} and will be probed further by near-future experiments \cite{Amendola:2016saw, *Abell:2009aa, *Levi:2013gra, *Aghamousa:2016zmz, *Spergel:2015sza, *Font-Ribera:2013rwa, *Jain:2015cpa}. This model also predicts neutrinoless double beta decay \cite{dell2016neutrinoless, *engel2017status, *vergados2016neutrinoless, *pas2015neutrinoless, *king2013power} with the invariant mass parameter $|m_{\beta \beta}| = 13.02\ \text{or}\ 25.21\ \text{meV}$, within an order of magnitude of the latest upper bound of $61-165\ \text{meV}$ measured by the KamLAND-Zen experiment \cite{gando2016search} and sensitive to several next-generation experiments \cite{Abgrall:2017syy, *Albert:2017hjq, *Wang:2015raa, *Wang:2015taa, *Andringa:2015tza, *Fischer:2018squ, *Jo:2017jod, *Chen:2016qcd}.} 


\blue{In this paper we expand the analysis of the asymmetric texture to investigate the generation of the baryon asymmetry \cite{Ignatiev:1978uf, *Yoshimura:1978ex, *Toussaint:1978br, *Weinberg:1979bt, *Yoshimura:1979gy, *Barr:1979ye, *Nanopoulos:1979gx, *Yildiz:1979gx} of the universe through leptogenesis \cite{Fukugita:1986hr}.} Baryon asymmetry is defined as the ratio of the net number of baryons to the number of photons: $\eta_B = (N_B - N_{\bar{B}})/N_{\gamma}$.
The abundance of matter over antimatter in the universe implies $\eta_B > 0$, as evidenced by the measurement from Cosmic Microwave Background (CMB) data \cite{Akrami:2018vks}:
\begin{align}
    \eta_B^{CMB} &= (6.12 \pm 0.04) \times 10^{-10}. \label{measuredB}
\end{align}
Sakharov identified three necessary conditions for successful generation of the baryon asymmetry \cite{Sakharov:1967dj}: (i) the existence of baryon number, $B$, violating elementary processes, (ii) violation of $C$ and $CP$, and (iii) a departure from thermal equilibrium. In leptogenesis, lepton asymmetry is generated from the $\cancel{C}$ and $\cancel{CP}$ out-of-equilibrium decays of the Majorana neutrinos into leptons and Higgs bosons. These decays violate the total lepton number $L$, which is partially converted into violation of the baryon number $B$ by $B-L$-preserving sphaleron processes \cite{Kuzmin:1985mm, *Khlebnikov:1988sr, *Harvey:1990qw}, fulfilling Sakharov's conditions.

We discuss leptogenesis in the so-called ``strong washout'' regime where only decays and inverse decays of the Majorana neutrinos describe generation of the asymmetry \cite{Buchmuller:2004nz, Buchmuller:2005eh, *Davidson:2008bu, *DiBari:2012fz}. We show that the low energy $\cancel{CP}$ phases of the model do not yield any high energy $CP$ asymmetry unless ``flavor effects'' \cite{Barbieri:1999ma, *Abada:2006fw, *Blanchet:2006be, *Vives:2005ra, *Nardi:2006fx, Abada:2006ea, Dev:2017trv, Samanta:2020tcl, *Samanta:2019yeg} are considered. The relevant density matrix equations are solved numerically for the non-hierarchical mass spectrum of the Majorana neutrinos. Successful leptogenesis occurs for Majorana masses of $\mc O(10^{11} - 10^{12})$ $\text{GeV}$ and constrains the parameter space of the model. 

The signs of the low energy leptonic $CP$ violation and the baryon asymmetry can, in general, be correlated \cite{Frampton:2002qc, *Achiman:2004qf, *Kaneta:2016gbq}. In the $SU(5) \times \mc T_{13}$ model the baryon asymmetry is generated by the single TBM phase whose sign was unresolved in the previous works \cite{Rahat:2018sgs, Perez:2020nqq}. We demonstrate that the final asymmetry is sensitive to this sign. The Dirac $\cancel{CP}$ phase $\delta_{CP}$ predicted in this model is $\pm 1.32 \pi$, compared to the latest PDG fit $\delta_{CP}^{PDG} = 1.36 \pm 0.17 \pi$ \cite{pdglive}. We identify the region of the parameter space which yields positive baryon asymmetry for the `correct' sign of $\delta_{CP}$.



The paper is organized as follows. In section \ref{sec:2}, we set up the lepton sector of the $SU(5) \times \mc T_{13}$ model presented in Refs.~\cite{Rahat:2018sgs,Perez:2019aqq,Perez:2020nqq} in a basis relevant for leptogenesis calculation. In section \ref{sec:3}, we briefly review thermal leptogenesis in the non-hierarchical mass spectrum of the Majorana neutrinos. In section \ref{sec:CP}, we discuss the relation between low energy $\cancel{CP}$ phases and high energy $CP$ asymmetry and show that leptogenesis is only viable in this model when flavor effects are taken into account.
Section \ref{sec:4} describes our results for the Majorana masses required for leptogenesis.
In section \ref{sec:5} we discuss how the sign of the TBM phase is correlated with the sign of the baryon asymmetry produced in leptogenesis and we conclude in section \ref{sec:6}.




\section{Lepton sector of the $SU(5) \times \mc T_{13}$ Model} \label{sec:2}
The ``asymmetric texture'' \cite{Rahat:2018sgs} is inspired by the $SU(5)$ Georgi-Jarlskog texture \cite{Georgi:1979df} with a $\overline{\g{45}}$ Higgs  coupling to the $(22)$ element of the down-quark and the charged-lepton Yukawa matrices $Y^{(-\frac{1}{3})}$ and $Y^{(-1)}$, respectively, and a $\gb{5}$ Higgs coupling elsewhere:
\begin{align}
\begin{aligned}
Y^{(\frac{2}{3})} \sim~ &\mathrm{diag}\ (\lambda^8, \lambda^4, 1), \\
Y^{(-{1\over3})} \sim 
\begin{pmatrix}
 b d \lambda ^4 & a \lambda ^3 & b \lambda ^3 \cr
 a \lambda ^3 & c \lambda ^2 & g \lambda ^2 \cr
 d \lambda  & g \lambda ^2 & 1 \end{pmatrix}
~~&\mathrm{ and~~~}
Y^{(-1)} \sim 
\begin{pmatrix}
 b d \lambda ^4 & a \lambda ^3 & d \lambda \cr
 a \lambda ^3 & -3c \lambda ^2 & g \lambda ^2 \cr
 b \lambda^3  & g \lambda ^2 & 1 \end{pmatrix}. \label{texture}
\end{aligned}
\end{align}
The $\mc O(1)$ prefactors $a = c = \frac{1}{3}$, $g = A$, $b = A \sqrt{\rho^2+\eta^2}$, $d = \frac{2}{3A}$ are determined in terms of the Wolfenstein parameters $A$, $\lambda $ $\rho$, and $\eta$. \blue{The asymmetry of $\mc O(\lambda)$ lies along the $(13)-(31)$ axis of $Y^{(-\frac{1}{3})}$ and $Y^{(-1)}$.} The up-quark Yukawa matrix $Y^{(\frac{2}{3})}$ is assumed to be diagonal. $SU(5)$ dictates $Y^{(-\frac{1}{3})}$ to be transpose of $Y^{(-1)}$ and the factor of $-3$ in the later comes from the vacuum expectation value of the $\overline{\g{45}}$ Higgs. 
The Yukawa matrices are unitarily diagonalized as $Y^{(q)} =$ $ \mc U^{(q)} \mc D^{(q)} \mc V^{(q)^\dagger}$, where $\mc U^{(-\frac{1}{3})} = \mc U_{CKM}$ and 
\begin{equation}
\mc U^{(-1)} =\left(
\begin{array}{ccc}
 1-\left(\frac{2}{9 A^2}+\frac{1}{18}\right) \lambda ^2 & \frac{\lambda }{3} & \frac{2 \lambda }{3 A} \\[0.5em]
 -\frac{\lambda }{3} & 1-\frac{\lambda ^2}{18} & A \lambda ^2 \\[0.5em]
 -\frac{2 \lambda }{3 A} & \left(-A-\frac{2}{9 A}\right) \lambda ^2 & 1-\frac{2 \lambda ^2}{9 A^2} \\
\end{array}
\right)+\mathcal{O}(\lambda^3). \label{ulep}
\end{equation}
\noindent Together with the complex-TBM seesaw mixing $\mc U_{seesaw} = \text{diag} (1,1,e^{i\delta})\ \mc U_{TBM}$, where $|\delta| \simeq 78^\circ$, this texture reproduces the GUT-scale mass ratios and the mixing angles of quarks and charged leptons and predicts Dirac and Majorana $CP$ violating phases in the lepton sector.

\blue{A straightforward explanation of the asymmetric term in $Y^{(-\frac{1}{3})}$ and $Y^{(-1)}$ requires an $SU(3)$-subgroup family symmetry with at least two different triplets. The smallest discrete group that fits the bill is $\mc T_{13}$  \cite{Perez:2019aqq}}. An $SU(5) \times \mc T_{13}$ model of effective interactions, where the $SU(5)$ matter fields transform as different triplets of $\mc T_{13}$ but the Higgs bosons are family singlets, explains the structure of the texture \cite{Perez:2019aqq} and the origin of the complex-TBM seesaw mixing \cite{Perez:2020nqq} through simple vacuum alignment of gauge-singlet family-triplet familons. The generic setup of three Majorana neutrinos appears to be in tension with the oscillation data. A minimal extension of the seesaw sector with a fourth Majorana neutrino resolves this and predicts normal ordering of the light neutrino masses.


\blue{The aim of this paper is to further investigate the seesaw sector of the model to see if the low energy $\cancel{CP}$ phases can explain the baryon asymmetry of the universe at high energies through leptogenesis.} We assume that both the gauge and the family symmetry are broken down to the Standard Model gauge group before this happens, so that the Majorana neutrinos decay into Standard Model leptons and Higgs. This implies that the mass of the Majorana neutrinos should be lower than $10^{16}$ $\text{GeV}$, the breaking scale of the gauge and family symmetry.

In the following subsections, we will briefly review the seesaw sector of the $SU(5) \times \mc T_{13}$ model and its breaking to the Standard Model gauge group. Then we will set up the relevant parameters in the appropriate basis for discussing leptogenesis in the subsequent sections.

 
\subsection{From $SU(5) \times \mc T_{13}$ to the Standard Model Gauge Group}
The seesaw Lagrangian of the $SU(5) \times \mc T_{13}$ model \cite{Perez:2020nqq} is given by
\begin{align}
    \mc L_{ss} &\supset  y_{\mc A} F \Lambda \bar{H}_{\g{5}} + y'_{\mc A} \bar{N} \overline{\Lambda} \varphi_{\mc A} + y_{\mc B} \bar{N} \bar{N} \varphi_{\mc B}   + M_\Lambda \overline{\Lambda} \Lambda + y_v' \bar{N}_4 \overline{\Lambda} \varphi_v + m \bar{N}_4 \bar{N}_4, \label{seesawlag}
\end{align}
where $y_X$ are dimensionless Yukawa couplings, $M_\Lambda$ is the mass of the heavy vectorlike messenger $\Lambda$ and $m$ is the mass of the fourth right handed neutrino $\bar{N}_4$. While we treat $m$ as an yet undetermined mass scale, it could originate from the vacuum expectation value (VEV) of a singlet familon and thus be related to the family symmetry breaking scale. 
The Lagrangian in Eq.~\eqref{seesawlag} has a $\mc Z_{12}$ `shaping' symmetry to prevent unwanted operators.\footnote{Ref.~\cite{Perez:2020nqq} also discusses a $\mc Z_{14}$ `shaping' symmetry for a slightly different particle content. In Appendix \ref{app:Z14} we show that this case does not yield successful leptogenesis for the simplest vacuum alignments of familons. More general cases do yield nonzero baryon asymmetry. It is beyond the scope of this paper and will be discussed in a future work.} Charged leptons reside in the field $F$ and the Majorana neutrinos in $\bar{N}$ and $\bar{N}_4$. The three familons $\varphi_{\mc A}$, $\varphi_{\mc B}$ and $\varphi_v$ have VEVs given by: 
\begin{align*}
\vev{\varphi_{\mc A}} &= \frac{M_{\Lambda}}{\vev{\bar{H}_{\g{5}}}} \sqrt{m_\nu b_1 b_2 b_3}\ (-b_2^{-1} e^{i\delta}, b_1^{-1}, b_3^{-1}),\\ \vev{\varphi_{\mc B}} &= (b_1, b_2, b_3), \\
\vev{\varphi_v} &= \frac{M_{\Lambda}}{\vev{\bar{H}_{\g{5}}}} \sqrt{mm_v'}\ (2,-1,e^{i\delta}),
\end{align*}
where $b_1, b_2, b_3, m \neq 0$. \blue{The vacuum alignments of $\varphi_{\mc A}$ and $\varphi_{\mc B}$ are related to each other, as required for the complex-TBM diagonalization of the seesaw matrix.}\footnote{This vacuum alignment relates the family symmetry breaking scale to the messenger scale. Suppose $b_1 \sim b_2 \sim b_3 \sim m \sim 10^{11}\ \text{GeV}$, a reasonable scale for family symmetry breaking and Majorana masses. Since $m_\nu \sim m_v' \sim 10^{-11}\ \text{GeV}$, cf. Eq.~\eqref{mnu}, the scale of $\vev{\varphi_{\mc A}}$ and $\vev{\varphi_v}$ would be similar to $\vev{\varphi_{\mc B}}$ if $M_\Lambda \sim 10^{13}\ \text{GeV}$.} The transformation properties of the relevant fields are given in Table \ref{table:model}.
\begin{table}[ht]\centering
\renewcommand\arraystretch{1.1}
\begin{tabularx}{\textwidth}{@{}l | Y Y Y Y Y Y Y Y @{}}
\toprule
     & $F$ & $\bar{N}$ & $\bar{N}_4$ & $\bar{H}_{\g{5}}$ & $\Lambda$  & $\varphi_{\mc A}$ & $\varphi_{\mc B}$ & $\varphi_v$ \\ 
\hline
$SU(5)$    & $\overline{ \g{5}}$ & $\g{1}$ & ${ \g{1}}$ & ${\g{5}}$ & $\g{1}$ & ${ \g{1}}$ & $\g{1}$ & $\g{1}$ \\ 
$\mc T_{13}$ &  $\g{3}_1$ & $\g{3}_2$ & $\g{1}$ & $\g{1}$ & $\gb{3}_1$  & $\gb{3}_2$ & $\g{3}_2$ & $\gb{3}_1$  \\ 
$\mathcal{Z}_{12}$    & $\g{\omega}$ & $\g{\omega^3}$ & $\g{1}$ & $\g{\omega^9}$ & $\g{\omega^2}$  & $\g{\omega^{11}}$ & $\g{\omega^6}$ & $\g{\omega^2}$   \\ 
\botrule
\end{tabularx} 
\caption{Charge assignments of matter, Higgs, messenger and familon fields in the seesaw sector. Here $\omega^{12} = 1$. The $\mc Z_{12}$ `shaping' symmetry is required to prevent unwanted tree-level operators.}
\label{table:model}
\end{table}

Using oscillation data, the parameters $m_{\nu}$ and $m_v'$ were determined in Ref.~\cite{Perez:2020nqq} as
\begin{align}
    |m_\nu| = 57.8\ \text{meV},  \quad |m_v'| = 5.03\ \text{or}\ 14.2\ \text{meV}. \label{mnu}
\end{align}
For our calculation, we will adopt $m_\nu = 57.8\ \text{meV}$ and $m_v' = 5.03\ \text{meV}$.  
%
%
\blue{This leaves four undetermined parameters: $b_1, b_2, b_3$ and $m$. In this paper we will discuss how these parameters are constrained when successful leptogenesis occurs.}




Integrating out the heavy messenger $\Lambda$ from the Lagrangian in Eq.~\eqref{seesawlag} gives the dimension-5 operators $\frac{1}{M_{\Lambda}} F \bar{N} \bar{H}_{\g{5}} \varphi_{\mc A}$ and $\frac{1}{M_{\Lambda}} F \bar{N}_4 \bar{H}_{\g{5}} \varphi_{v}$. These operators yield the Dirac Yukawa matrix $Y^{(0)}$ when the familon $\varphi_{\mc A}$ and $\varphi_v$ develop nonzero VEVs spontaneously breaking the $\mc T_{13} \times \mc Z_{12}$ symmetry  \cite{Perez:2020nqq}:
\begin{align}
    Y^{(0)} &\equiv \frac{\sqrt{b_1 b_2 b_3 m_\nu}}{\vev{\bar{H}_\g{5}}} \left(
\begin{array}{cccc}
 0 & b_3^{-1} & 0 & 2\sqrt{\frac{mm_v'}{b_1 b_2 b_3 m_\nu}}  \\[0.5em]
 b_1^{-1} & 0 & 0 & -\sqrt{\frac{mm_v'}{b_1 b_2 b_3 m_\nu}} \\[0.5em]
 0 & 0 & -e^{i \delta}b_2^{-1} & e^{i \delta} \sqrt{\frac{mm_v'}{b_1 b_2 b_3 m_\nu}} \\
\end{array}
\right). \label{Y0mat}
\end{align}
The effective operator $F \bar{N} \bar{H}_{\g{5}}$ further gives rise to the interaction $\ell_\alpha^\dagger H^* \bar{N}_i$ when the $SU(5)$ symmetry is broken down to the Standard Model gauge group and generates the decays:
\begin{align}
    \bar{N}_i \rightarrow \ell_\alpha^\dagger + H^*, \qquad i = 1, 2, 3, 4; \quad \alpha = e, \mu, \tau.
\end{align}



The $4 \times 4$ Majorana mass matrix gets contribution from the VEV of the familon $\varphi_{\mc B}$  and can be expressed as \cite{Perez:2020nqq}
\begin{align}
    \mathcal{M} &\equiv \left(
\begin{array}{cccc}
 0 & b_2 & b_3 & 0 \\
 b_2 & 0 & b_1 & 0 \\
 b_3 & b_1 & 0 & 0 \\
 0 & 0 & 0 & m \\
\end{array}
\right). \label{Mmat}
\end{align}
It is a complex symmetric matrix and its Takagi factorization \cite{horn2012matrix} yields
\begin{align}
    \mc M = \mc U_m\ \mc D_m\ \mc U_m^T. \label{majorana}
\end{align}
Here $\mc D_m = \text{diag}(M_1, M_2, M_3, M_4)$ is the diagonal mass matrix with the positive square root of real eigenvalues of $\mc M \mc M^\dagger$ and $\mc U_m $ is the unitary matrix containing the corresponding eigenvectors of  $\mc M \mc M^\dagger$.\footnote{\blue{The matrices in Eqs.~\eqref{Y0mat} and \eqref{Mmat} are  valid at the grand unified scale ($\sim 10^{16}\ \text{GeV}$). For simplicity we neglect the effects of their running and assume that they are valid at the Majorana mass scale ($\sim 10^{11}$ $\text{GeV}$) too. See Ref.~\cite{Antusch:2005gp} for more discussion on the effect of running seesaw parameters on leptogenesis.}}   

\subsection{Rotating to the \emph{Weak} Basis}
In leptogenesis we usually work in the so-called \emph{weak} basis, where the charged-lepton Yukawa matrix and the right-handed Majorana matrix are diagonal with real, positive entries \cite{Zhang:2015qia}. 
After spontaneous breaking of the $SU(5) \times \mc T_{13} \times \mc Z_{12}$ symmetry, the relevant terms in the Lagrangian are
\begin{align}
    \mc L &\supset \ell^\dagger Y^{(-1)} \bar{e}H + \ell^\dagger Y^{(0)}\bar{N}H^* + \bar{N}^T \mc M \bar{N} \nonumber \\
    &= \ell^\dagger \mc U^{(-1)} \mc D^{(-1)} \mc V^{(-1)\dagger} \bar{e}H + \ell^\dagger Y^{(0)}\bar{N}H^* + \bar{N}^T \mc U_m \mc D_m \mc U_m^T \bar{N}. \label{lagrangian}
\end{align}
Redefining the fields $\ell \rightarrow \mc U^{(-1)}\ell$, $\bar{e} \rightarrow \mc V^{(-1)}\bar{e}$, and $\bar{N} \rightarrow \mc U_m^* \bar{N}$, it becomes
\begin{align}
    \mc L \supset \ell^\dagger \mc D^{(-1)\dagger} \bar{e}H + \ell^\dagger \mc U^{(-1)\dagger} Y^{(0)} \mc U_m^* \bar{N} H^* + \bar{N}^T \mc D_m \bar{N},
\end{align}
and we identify the light neutrino Yukawa matrix:
\begin{align}
    Y_\nu = \mc U^{(-1)\dagger} Y^{(0)} \mc U_m^*. \label{first}
\end{align}
%
$Y_\nu$ serves as a key input for leptogenesis.
\section{Thermal leptogenesis in the non-hierarchical mass spectrum} \label{sec:3}

In this section we will briefly review the formalism of thermal leptogenesis relevant for our discussion later.
Majorana neutrinos are produced in the early universe from Yukawa interactions of leptons and Higgs bosons in a thermal bath right below the very high reheating temperature $T_{RH} \lesssim 10^{15}\ \text{GeV}$ \cite{Fukugita:2003en, Khlopov:1984pf, *balestra1984annihilation, *khlopov1994nonequilibrium, *Khlopov:2004tn}. Any pre-existing asymmetry is completely diluted by inflation and the Majorana neutrinos are in thermal equilibrium. As the temperature falls below their mass, their overabundance above the equilibrium density prompts decays into leptons (with a decay width $\Gamma_{i\alpha}$) or into antileptons (with a decay width $\overline{\Gamma}_{i\alpha}$). These  $\cancel{L}$, $\cancel{C}$ and $\cancel{CP}$ processes go out of equilibrium as the decay rate becomes smaller than the expansion rate of the universe.
At $100 \ll T\ (\text{GeV}) \ll 10^{12}$, sphaleron processes, which violate both $B$ and $L$ but conserve $B-L$, are in equilibrium and convert part of the generated lepton asymmetry to the baryon asymmetry \cite{Kuzmin:1985mm, *Khlebnikov:1988sr, *Harvey:1990qw}.

Leptogenesis is a battle between decays and inverse decays of the Majorana neutrinos. The minimal scenario involves a hierarchical mass spectrum, where the asymmetry generated by the decay of the heavier Majorana neutrinos is washed out as the temperature comes down to the scale of the lightest mass and the final baryon asymmetry is generated entirely from its decay. Such scenarios appear, for example, in $SO(10)$-inspired models \cite{VelascoSevilla:2003gd, *DiBari:2008mp, *DiBari:2014eya, *Chen:2014wiw}, where the Majorana masses follow the hierarchy of the up-quark masses with a suppression of $\mc O(\lambda^4)$ between families. For a non-hierarchical mass spectrum, however, one must consider the decay of all Majorana neutrinos, since the asymmetry generated by the decay of the heavier ones are not completely washed out \cite{Blanchet:2006dq}.

All flavors of the charged leptons in the decay product can be considered identical as long as the lightest Majorana neutrino mass is far above $10^{12}$ $\text{GeV}$, a scenario known as ``unflavored leptogenesis''. Flavor plays an important role for smaller mass scales and can enhance the final asymmetry significantly \cite{Abada:2006ea, Dev:2017trv}. 

In the following, we will discuss both of these cases and express the relevant equations in terms of the seesaw parameters $Y_\nu$ and $M_i$. 

\subsection{``Flavored'' leptogenesis}

The evolution of number density of the Majorana neutrino $N_{N_i}$ is kinematically described by the following equation \cite{luty1992baryogenesis, Blanchet:2006dq}:
\begin{align}
    \frac{dN_{N_i}}{dz} &= -(D_i + S_i) (N_{N_i} - N_{N_i}^{eq}), \label{eq:Nni1}
\end{align}
where $z \equiv M_{min}/T$ and $M_{min} \equiv \text{min} (M_i)$. 

Introducing the notation $x_i \equiv M_i^2 / M_{min}^2$ and $z_i \equiv z\sqrt{x_i}$,
the \emph{equilibrium number density} can be expressed in terms of the modified Bessel functions of the second kind \cite{Buchmuller:2004nz}:
\begin{align}
    N_{N_i}^{eq} (z_i) &= \frac{1}{2} z_i^2 \mc K_2(z_i),
\end{align}
so that $N_{N_i}^{eq} (z_i \ll 1) = 1$. The \emph{decay factor} $D_i$ is given by \cite{Kolb:1979qa, Buchmuller:2004nz}
\begin{align}
    D_i \equiv \frac{\Gamma_{D,i}}{H(z_i)\ z_i} = K_i x_i z \frac{\mc K_1(z_i)}{\mc K_2(z_i)} \label{eq:Di1}, 
\end{align}
where $\Gamma_{D,i} \equiv \Gamma_i + \bar{\Gamma}_i$ is the \emph{total decay rate} and $H(z_i)$ is the Hubble expansion rate. The \emph{decay parameter} $K_i$ is given by  \cite{luty1992baryogenesis, Kolb:1990vq}:
\begin{align}
    K_i \equiv \frac{\widetilde{\Gamma}_{D,i} }{H(z_i = 1)} = \frac{(Y_\nu^\dagger Y_\nu)_{ii}}{M_i m_*}, \label{Ki}
\end{align}
where $\widetilde{\Gamma}_{D,i} \equiv \Gamma_{D,i}(z_i = \infty)$ and $m_* \simeq 1.07\ \text{meV}$ is the \emph{effective neutrino mass} \cite{Plumacher:1996kc}. If we limit our discussion to the scenario when $K_{i} \gg 1$, i.e., the so-called ``strong washout'' region, the dynamics can be explained well by considering only decays and inverse decays \cite{Buchmuller:2004nz} and the $\Delta L = 1$ \emph{scattering term} $S_i$ can be neglected.

Flavor effects become significant in models where unflavored leptogenesis is not viable and/or scenarios where the mass of the Majorana neutrinos are below $10^{12}\ \text{GeV}$. For $10^9 \ll T\ (\text{GeV}) \ll 10^{12}$, the tau leptons can decohere and the dynamics of leptogenesis can be described in terms of two-flavor approximate Boltzmann equations. For $T \ll 10^9\ \text{GeV}$, the muons also decohere and the evolution of $B-L$ asymmetry can be tracked with three-flavor approximate Boltzmann equations. The more general description of the dynamics can be achieved with the density matrix formalism, where one considers the flavor space as a $3\times 3$ matrix and accounts not only for the three flavors but also for the transition between them. 


The most general form of the density matrix equations for the evolution of the $B-L$ asymmetry is given by \cite{Blanchet:2011xq}
\begin{align}
    \frac{dN_{\alpha \beta}}{dz} &= \sum_{i} \varepsilon_{\alpha \beta}^{(i)} D_i (N_{N_i} - N_{N_i}^{eq}) -\frac{1}{2} \sum_i W_i \{P^{0(i)},N\}_{\alpha \beta} \nonumber\\
    &- \frac{\text{Im}(\Lambda_\tau)}{Hz}\left[\left(
\begin{array}{ccc}
 1 & 0 & 0 \\
 0 & 0 & 0 \\
 0 & 0 & 0 \\
\end{array}
\right),\left[\left(
\begin{array}{ccc}
 1 & 0 & 0 \\
 0 & 0 & 0 \\
 0 & 0 & 0 \\
\end{array}
\right),N \right]\right]_{\alpha \beta} \nonumber\\
 &- \frac{\text{Im}(\Lambda_\mu)}{Hz}\left[\left(
\begin{array}{ccc}
 0 & 0 & 0 \\
 0 & 1 & 0 \\
 0 & 0 & 0 \\
\end{array}
\right),\left[\left(
\begin{array}{ccc}
 0 & 0 & 0 \\
 0 & 1 & 0 \\
 0 & 0 & 0 \\
\end{array}
\right),N \right]\right]_{\alpha \beta},   \label{eq:NBLdm}
\end{align}
where $\Lambda_{\alpha}$ is the self-energy of $\alpha$-flavored leptons. The \emph{thermal widths} are given by the imaginary part of the self-energy correction to the lepton propagator in the plasma, and can be expressed as \cite{Moffat:2018wke}
\begin{align}
    \frac{\text{Im}(\Lambda_\tau)}{Hz} &= 4.66 \times 10^{-8} \frac{M_{Pl}}{M_{min}}, \\
    \frac{\text{Im}(\Lambda_\mu)}{Hz} &= 1.69 \times 10^{-10} \frac{M_{Pl}}{M_{min}},
\end{align}
where $M_{Pl} = 1.22 \times 10^{19}\ \text{GeV}$ is the Planck mass. The projection matrices $P^{0(i)}$ are defined as 
\begin{align}
    P^{0(i)}_{\alpha \beta} \equiv \frac{(Y_\nu^*)_{\alpha i} (Y_\nu)_{\beta i}}{(Y_\nu^\dagger Y_\nu)_{ii}}
\end{align}
and describe how a given flavor of lepton is washed out. The $CP$ asymmetry matrix $\varepsilon^{(i)}$ denotes the decay asymmetry generated by $N_i$ and its elements are perturbatively calculated from the interference of the tree-level with the one-loop and the self-energy diagrams when $|M_j - M_i|/M_i \gg \text{max}[(Y_\nu^\dagger Y_\nu)_{ij}]/(16\pi^2 )$ 
\cite{Covi:1996wh, Zhang:2015qia}:
\begin{align}
    \varepsilon^{(i)}_{\alpha \beta} &= \frac{1}{16 \pi (Y^\dagger_\nu Y_\nu)_{ii}} \sum_{j \neq i} \left\{i \left[ (Y_\nu)_{\alpha i} (Y_\nu^*)_{\beta j} (Y^\dagger_\nu Y_\nu)_{ji} - (Y_\nu^*)_{\beta i} (Y_\nu)_{\alpha j} (Y^\dagger_\nu Y_\nu)_{ij} \right] \zeta\left( \frac{x_j}{x_i} \right) \right. \nonumber \\
    &+ \left. i \left[ (Y_\nu)_{\alpha i} (Y_\nu^*)_{\beta j} (Y^\dagger_\nu Y_\nu)_{ij} - (Y_\nu^*)_{\beta i} (Y_\nu)_{\alpha j} (Y^\dagger_\nu Y_\nu)_{ji} \right] \xi\left( \frac{x_j}{x_i} \right)
    \right\}, \label{epsidm}
\end{align}
where the loop factors are given by
\begin{align}
    \xi(x) &= \sqrt{x} \left( (1+x) \log{\left(\frac{1+x}{x}\right)} + \frac{1}{x-1} - 1 \right), \qquad \text{and}\qquad
    \zeta(x) = \frac{1}{x-1},
\end{align}
which blow up if there is a mass degeneracy $x_i = x_j$. Although exact degeneracies cannot generate $CP$ asymmetry, nearly degenerate masses can significantly enhance the $CP$ asymmetry leading to a scenario known as ``resonant leptogenesis'' \cite{Flanz:1996fb, *Covi:1996fm, *Pilaftsis:1997dr, *Pilaftsis:1997jf, *Pilaftsis:1998ct, *Pilaftsis:2003gt, *Pilaftsis:2005rv, *Anisimov:2005hr}.

The \emph{washout term} $W_i$ represents the washout of the generated asymmetry for each Majorana neutrino. Subtracting the resonant contribution from $\Delta L = 2$ processes ($\ell_\alpha + H^* \leftrightarrow \bar{\ell}_\alpha + H$) to the inverse decays, it is given by \cite{Blanchet:2006dq}
\begin{align}
    W_i \equiv W_i^{ID}(z)  = \frac{1}{4} K_i \sqrt{x_i} \mc K_1(z_i) z_i^3.
\end{align}

Solving the system of equations \eqref{eq:Nni1} and \eqref{eq:NBLdm} yields the flavor-dependent asymmetry $N_{\alpha \beta}$, which is, in general, a $3 \times 3$ matrix.  The total lepton asymmetry is the trace of this matrix: 
\begin{align}
    N_{B-L} \equiv \sum_{\alpha = e, \mu, \tau} N_{\alpha \alpha}
\end{align}
and its final value $N_{B-L}^f$ is related to the baryon asymmetry by
\begin{align}
    \eta_B = a_{sph} \frac{N_{B-L}^f}{N_\gamma^{rec}} \simeq 0.96 \times 10^{-2} N_{B-L}^f, \label{formulaB}
\end{align}
where the sphaleron conversion coefficient is $a_{sph} = 28/79$ \cite{Kuzmin:1985mm, *Khlebnikov:1988sr, *Harvey:1990qw} and the baryon-to-photon number ratio at recombination is $N^{rec}_\gamma \simeq 37$ \cite{Blanchet:2006dq}. Successful leptogenesis requires $\eta_B$ to match the measured value in Eq.~\eqref{measuredB}. 

In three-flavor approximate Boltzmann equations, the off-diagonal components of $N_{\alpha \beta}$ are ignored. The evolution of the $B-L$ asymmetry is
split into individual equations for each flavor $\alpha = e, \mu, \tau$ \cite{Barbieri:1999ma, *Abada:2006fw, *Blanchet:2006be, *Vives:2005ra, *Nardi:2006fx}:
\begin{align}
    \frac{dN_{\alpha\alpha}}{dz} = -\sum_{i} \varepsilon_{\alpha \alpha}^{(i)} D_i (N_{N_i} - N_{N_i}^{eq}) - N_{\alpha \alpha} \sum_i P^{0(i)}_{\alpha \alpha}  W_i. \label{eq:NBLflav}
\end{align} 

\subsection{``Unflavored'' leptogenesis}
All flavor-dependent parameters are summed over the flavor index $\alpha$ in ``unflavored'' leptogenesis. This yields the flavor-independent $CP$-asymmetry parameter \cite{luty1992baryogenesis, Blanchet:2006dq}:
\begin{align}
    \varepsilon^{(i)} &\equiv \frac{\Gamma_i - \bar{\Gamma}_i}{\Gamma_i + \bar{\Gamma}_i} = \frac{1}{8\pi }\sum_{j\neq i} \frac{\text{Im}\left[\left(({Y_\nu^\dagger Y_\nu})_{ij}\right)^2\right]}{({Y_\nu^\dagger Y_\nu})_{ii}}\ \xi\left(\frac{x_j}{x_i}\right). \label{epsiloni}
\end{align}

Eq.~\eqref{eq:Nni1} still represents the evolution of number densities of Majorana neutrinos in the strong washout region. The flavor-independent $B-L$ asymmetry is described by the following Boltzmann equation \cite{luty1992baryogenesis, Blanchet:2006dq}:
\begin{align}
    \frac{dN_{B-L}}{dz} &= -\sum_i \varepsilon^{(i)} D_i (N_{N_i} - N_{N_i}^{eq}) - N_{B-L} \sum_i W_i^{ID}. \label{eq:NBL}
\end{align}

Eqs.~\eqref{eq:Nni1} and \eqref{eq:NBL} can be solved as coupled first order differential equations and their solution yields  $N_{B-L}$ in the unflavored case.

\section{Relating low energy  $CP$ violation to high energy $CP$ asymmetry} \label{sec:CP}
The relation between low energy $CP$ violation in the PMNS matrix and high energy $CP$ violation required for leptogenesis has been discussed extensively in literature \cite{Berger:1999bg, *Goldberg:1999hp, *Falcone:2001im, *Ellis:2002eh, *Davidson:2002em, *Branco:2002kt, *King:2002qh, *Endoh:2002wm, *Branco:2002xf, *Pascoli:2003uh}. In general, the existence of $\cancel{CP}$ phases in the PMNS matrix do not guarantee $CP$ asymmetry in unflavored leptogenesis. However, barring accidental cancellations, observation of low energy  $CP$ violation necessarily implies generation of the baryon asymmetry in flavored leptogenesis \cite{Pascoli:2006ie, *Branco:2006ce, *Mohapatra:2006se, *Pascoli:2006ci, *Uhlig:2006xf}. 

In the asymmetric texture, the only source for both Dirac and Majorana $CP$ violation is the TBM phase $\delta$,\footnote{ \blue{To clarify, it is related to but not the same as the Dirac phase $\delta_{CP}$ in the PMNS matrix.}} appearing in the matrix $\text{diag}(1,1,e^{i\delta})$ multiplying the real TBM matrix from the left. In this section we will argue that this particular placement of the phase results in vanishing $CP$ asymmetry in the unflavored case. 

The seesaw matrix is given by
\begin{align}
    \mc S &\equiv Y^{(0)} \mc M^{-1} Y^{(0)T}  \nonumber\\
    &= \left[ \mc D_m^{-1/2} \mc U_m^\dagger Y^{(0)T} \right]^T\ \left[ \mc D_m^{-1/2} \mc U_m^\dagger Y^{(0)T} \right], \label{seesawgen}
\end{align}
where $\mc D_m^{-1/2} \equiv \text{diag}(M_1^{-1/2}, M_2^{-1/2}, M_3^{-1/2}, M_4^{-1/2})$ is a diagonal matrix with all positive entries. Diagonalization of the seesaw matrix by the complex-TBM mixing implies
\begin{align}
    \mc S &= \text{diag} (1,1,e^{i\delta})\ \mc U_{TBM}\ \mc D_{\nu}\ \mc U_{TBM}^T\ \text{diag} (1,1,e^{i\delta}) \nonumber \\
    &= \left[\mc D_\nu^{1/2} \mc U_{TBM}^T\ \text{diag} (1,1,e^{i\delta})\right]^T \left[\mc D_\nu^{1/2} \mc U_{TBM}^T\ \text{diag} (1,1,e^{i\delta})\right], \label{TBMdiag}
\end{align}
where $\mc D_\nu^{1/2} \equiv \text{diag}(m_1^{1/2}, m_2^{1/2}, m_3^{1/2})$. In general the entries in $\mc D_\nu$ can be either positive or negative. Comparing Eqs.~\eqref{seesawgen} and \eqref{TBMdiag}, we find that $\mc D_m^{-1/2} \mc U_m^\dagger Y^{(0)T}$ has the following form:
\begin{align}
   \mc D_m^{-1/2} \mc U_m^\dagger Y^{(0)T} \equiv P W \text{diag} (1,1,e^{i\delta}), \label{seesawform}
\end{align}
where $W$ is a real matrix and $P$ is a diagonal phase matrix with entries either $1$ or $i$ (so that $P^T P =$ $ \text{diag}(\pm 1, \pm 1, \pm 1, \pm 1)$).

It is useful to define an orthogonal matrix $R$ in the Casas-Ibarra parametrization \cite{Casas:2001sr} to relate the low energy  parameters to the high energy $CP$ asymmetry:
\begin{align}
    R \equiv \mc D_m^{-1/2} \mc U_m^\dagger Y^{(0)T} \text{diag}(1,1,e^{-i\delta}) \mc U_{TBM}  \mc D_{\nu}^{-1/2},
\end{align}
where $R$ is complex in general. 
Then, from Eq.~\eqref{seesawform}, 
\begin{align}
    R = PW\mc U_{TBM}  \mc D_{\nu}^{-1/2}. \label{Rform}
\end{align}
In this parametrization, the neutrino Dirac Yukawa matrix can be written as, cf. Eq.~\eqref{first}:
\begin{align}
    Y_\nu &= \mc U_{PMNS} \mc D_\nu^{1/2} R^T \mc D_m^{1/2}, \label{YnuCI}
\end{align}
where $\mc U_{PMNS} = \mc U^{(-1)\dagger} \text{diag}(1,1,e^{i\delta})\mc U_{TBM}$, so that
\begin{align}
    Y_\nu^\dagger Y_\nu =  P^*\ \left(\mc D_m^{1/2}W  W^T \mc D_m^{1/2}\right)\ P. \label{YYCI}
\end{align}


The relation between low energy  $\cancel{CP}$ phases and high energy $CP$ asymmetry is evident from Eqs.~\eqref{YnuCI} and \eqref{YYCI}. $CP$ asymmetry in unflavored leptogenesis depends on ${\text{Im} \left[ (Y_\nu^\dagger Y_\nu)_{ij}^2 \right]}/{( Y_\nu^\dagger Y_\nu)_{ii}}$ for $j\neq i$, cf. Eq.~\eqref{epsiloni}.
From Eq.~\eqref{YYCI}, the diagonal elements of $ Y_\nu^\dagger Y_\nu$ are real and the off-diagonal elements are either real or purely imaginary. Hence the $CP$-asymmetry parameter vanishes and the low energy  $\cancel{CP}$ phases do not result in unflavored leptogenesis.

However, from Eq.~\eqref{epsidm}, the $CP$-asymmetry parameter in the density matrix formalism depends on $ ({Y_\nu^*})_{\alpha i} ({Y_\nu})_{\beta j} ({Y_\nu^\dagger Y_\nu})_{ji}$ and $({Y_\nu^*})_{\beta i} ({Y_\nu})_{\alpha j} ({Y_\nu^\dagger Y_\nu})_{ij}$ for $j \neq i$.
The $\cancel{CP}$ phases in the PMNS matrix do not vanish in $(Y_\nu^*)_{\alpha i} (Y_{\nu})_{\beta j}$, cf. Eq.~\eqref{YnuCI}, in general, and the $CP$-asymmetry parameter is nonzero.

\section{Flavored Leptogenesis in the $SU(5) \times \mc T_{13}$ Model} \label{sec:4}
In this section we employ the formalism developed so far to calculate the baryon asymmetry in the $SU(5) \times \mc T_{13}$ model through flavored leptogenesis. Since the mass scale of the right handed neutrinos is unknown at this level, we use the more general density matrix formalism instead of the three-flavor approximate Boltzmann equations. Matching the calculated baryon asymmetry to the observed value constrains the undetermined model parameters $b_1$, $b_2$, $b_3$ and $m$.

The predictions for the light neutrino masses and neutrinoless double beta decay in this model do not depend on the particular value of $b_1, b_2, b_3$ except that they must be nonzero \cite{Perez:2020nqq}. However, in the spirit of simplicity in vacuum alignments of the familons in the electroweak sector of the model \cite{Perez:2019aqq}, we are motivated to set $b_1, b_2$ and $b_3$ to be of the same order and consider two cases: (i) $(b_1, b_2, b_3) \equiv b(1, f, 1)$, and (ii) $(b_1, b_2, b_3) \equiv b(f, f, 1)$,  where $f \neq 1$ is an $\mc O(1)$ prefactor.\footnote{The case for $f=1$ does not yield nonzero baryon asymmetry, as discussed in Appendix \ref{app:feq1}.} We discuss flavored leptogenesis in both of these cases below.

\subsection{Case 1: $(b_1, b_2, b_3) \equiv b(1, f, 1)$}
The Dirac Yukawa matrix in this case becomes:
\begin{align}
    Y^{(0)} &= \frac{\sqrt{b f m_\nu}}{v} \left(
\begin{array}{cccc}
 0 & 1 & 0 & 2\beta   \\
 1 & 0 & 0 & -\beta  \\
 0 & 0 & -f^{-1} e^{i \delta} & \beta e^{i \delta}  \\
\end{array}
\right), \label{Y0}
\end{align}
where $\beta \equiv \sqrt{\frac{a m_v'}{f m_\nu }}$, $a \equiv \frac{m}{b}$ and $v = 174\ \text{GeV}$ is the Higgs VEV.
The Majorana matrix is given by
\begin{align}
    \mc M &=  b\left(
\begin{array}{cccc}
 0 & f & 1 & 0 \\
 f & 0 & 1 & 0 \\
 1 & 1 & 0 & 0 \\
 0 & 0 & 0 & a \\
\end{array}
\right). \label{Majo}
\end{align}
Its Takagi factorization, cf. Eq.~\eqref{majorana}, yields
\begin{align}
\begin{split}
    M_1 &= bf, \quad M_2 = \frac{b}{2} \left(\sqrt{f^2+8}-f\right),\quad
    M_3 = \frac{b}{2} \left(\sqrt{f^2+8}+f\right),\ \quad M_4 = a b, \label{Mmass}
\end{split}
\end{align}
and
\begin{align}
    \mc U_m = \left(
\begin{array}{cccc}
 -\frac{i}{\sqrt{2}} & \frac{-i}{2} \sqrt{1-\frac{f}{\sqrt{f^2+8}}} & \frac{1}{2} \sqrt{1+\frac{f}{\sqrt{f^2+8}}} & 0 \\ [1.5em]
 \frac{i}{\sqrt{2}} & \frac{-i}{2}  \sqrt{1-\frac{f}{\sqrt{f^2+8}}} & \frac{1}{2} \sqrt{1+\frac{f}{\sqrt{f^2+8}}} & 0 \\[1.5em]
 0 & \frac{i}{\sqrt{2}} \sqrt{1+\frac{f}{\sqrt{f^2+8}}} & \frac{1}{\sqrt{2}} \sqrt{1-\frac{f}{\sqrt{f^2+8}}} & 0 \\[1em]
 0 & 0 & 0 & 1 \\
\end{array}
\right). \label{Unitm}
\end{align}

For simplicity, we will limit our discussion to non-resonant thermal leptogenesis where the Majorana neutrino masses are required to be away from degeneracy. In Eq.~\eqref{Mmass}, $M_1$ and $M_2$ are degenerate for $f=1$, which justifies our assumption $f\neq 1$. $f$ lifts the degeneracy and makes leptogenesis viable. 

\begin{figure}[!ht]
    \centering
    \includegraphics[width=0.48\textwidth]{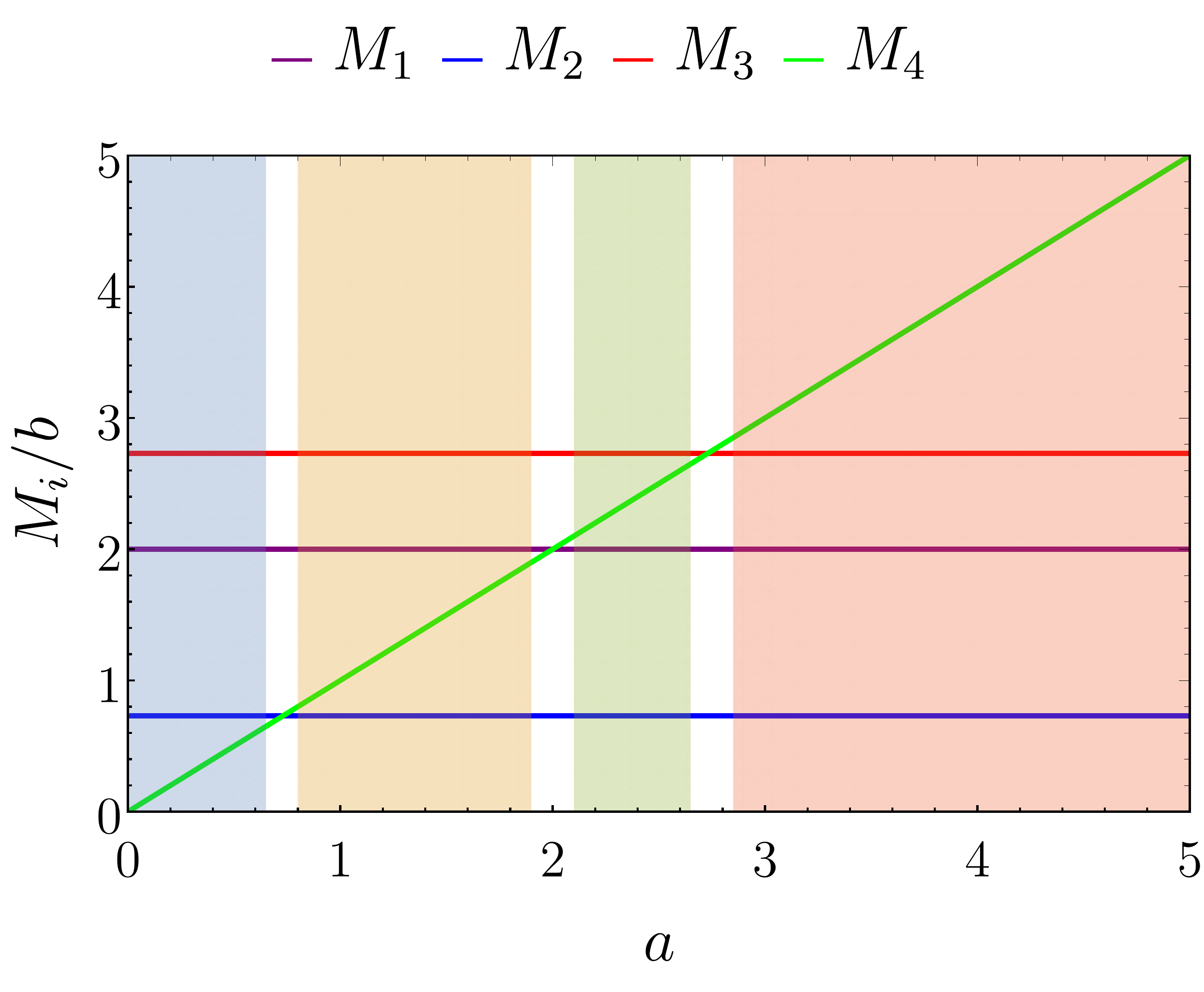}
\caption{Majorana neutrino mass spectrum for Case 1: $(b_1,b_2,b_3) \equiv (b(1,f,1)$. $M_1$, $M_2$ and $M_3$ do not depend on $a$. $M_4$ is degenerate with $M_1$, $M_2$ and $M_3$ for $a \simeq 0.73, 2$ and $2.73$, respectively, setting $f=2$. The parameter space can be divided into four regions to avoid near-degeneracies. } 
\label{fig:mass}
\end{figure}

For concreteness, we will set $f=2$ for the remainder of our discussion whenever a numerical value is required.\footnote{We require $f \sim \mc O(1)$ to avoid hierarchy among components of the VEVs of the familons, inspired from the VEVs of the electroweak familons of the model presented in Ref.~\cite{Perez:2019aqq}. We have verified that the final results relevant for leptogenesis are in the same order of magnitude as long as $f\sim \mc O(1)$.} This leaves us with two undetermined parameters $b$ and $a$, and yields the following mass spectrum:
\begin{align}
    \frac{M_1}{b} = 2,\quad \frac{M_2}{b} = \sqrt{3}-1, \quad \frac{M_3}{b} = \sqrt{3}+1, \quad \frac{M_4}{b} = a. \label{Mmass_model1}
\end{align}
Since only $M_4$ depends on $a$, it can be degenerate with $M_1$, $M_2$ and $M_3$ for $a = 2$, $\sqrt{3}-1 \simeq 0.73$ and $\sqrt{3}+1 \simeq 2.73$, respectively, as shown in Figure \ref{fig:mass}. To avoid resonant enhancement near degeneracies we split the parameter space into four regions: (i) $0.1 \leq a \leq 0.65$, (ii) $0.8 \leq a \leq 1.9$, (iii) $2.1 \leq a \leq 2.65$, and (iv) $a \geq 2.85$, shown in Figure \ref{fig:mass}. These regions represent particular mass ordering of the Majorana neutrinos. For example, region (ii) corresponds to $M_2 < M_4 < M_1 < M_3$. 
%




\textcolor{black}{We assume that there is no asymmetry present in any flavor in the universe before the decay of the Majorana neutrinos occur: $N_{\Delta \alpha}(z = 0) = 0$, and the reheating temperature of inflation is sufficiently higher than the mass of the heaviest Majorana neutrino, so that the asymmetry generated by the heavier ones is not washed out prior to the decay of the lightest one.}  


Although a more accurate picture of leptogenesis is depicted by the non-equilibrium thermal field theory approach \cite{Moffat:2018wke}, the density matrix equations discussed before are accurate as long as the strong washout condition $K_i \gg 1$ holds. An explicit calculation yields
\begin{align}
\begin{split}
    K_1 &= \frac{m_\nu}{ m_*} \simeq 54.0,\quad\ \ K_2 = \frac{m_\nu}{m_*} \frac{f \left(1-f^2\right)+\left(f^2+1\right) \sqrt{f^2+8}}{f \sqrt{f^2+8} \left(\sqrt{f^2+8}-f\right)} \simeq 60.3, \\
    K_4 &= \frac{6m_v'}{m_*} \simeq 28.2, \quad K_3 = \frac{m_\nu}{m_*} \frac{f \left(f^2-1\right)+\left(f^2+1\right) \sqrt{f^2+8}}{f \sqrt{f^2+8} \left(\sqrt{f^2+8}+f\right)} \simeq 33.3,
\end{split}
\end{align}
justifying this.

\begin{figure}[!ht] 
    \centering
    \subfloat[$a = 0.3$ and $f = 2$]{
        \includegraphics[width=0.48\textwidth]{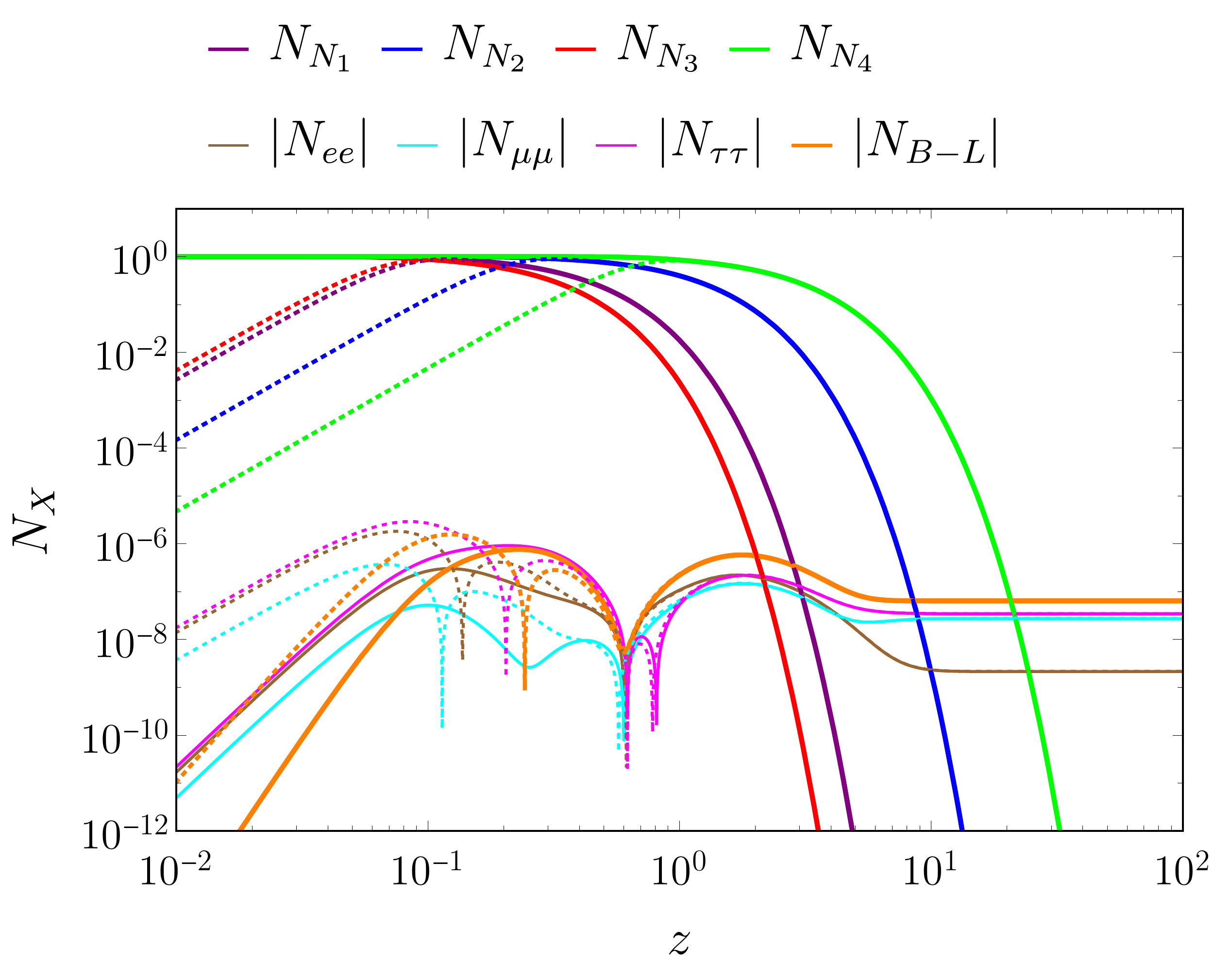}
    }
    \subfloat[$a = 1.4$ and $f = 2$]{
        \includegraphics[width=0.48\textwidth]{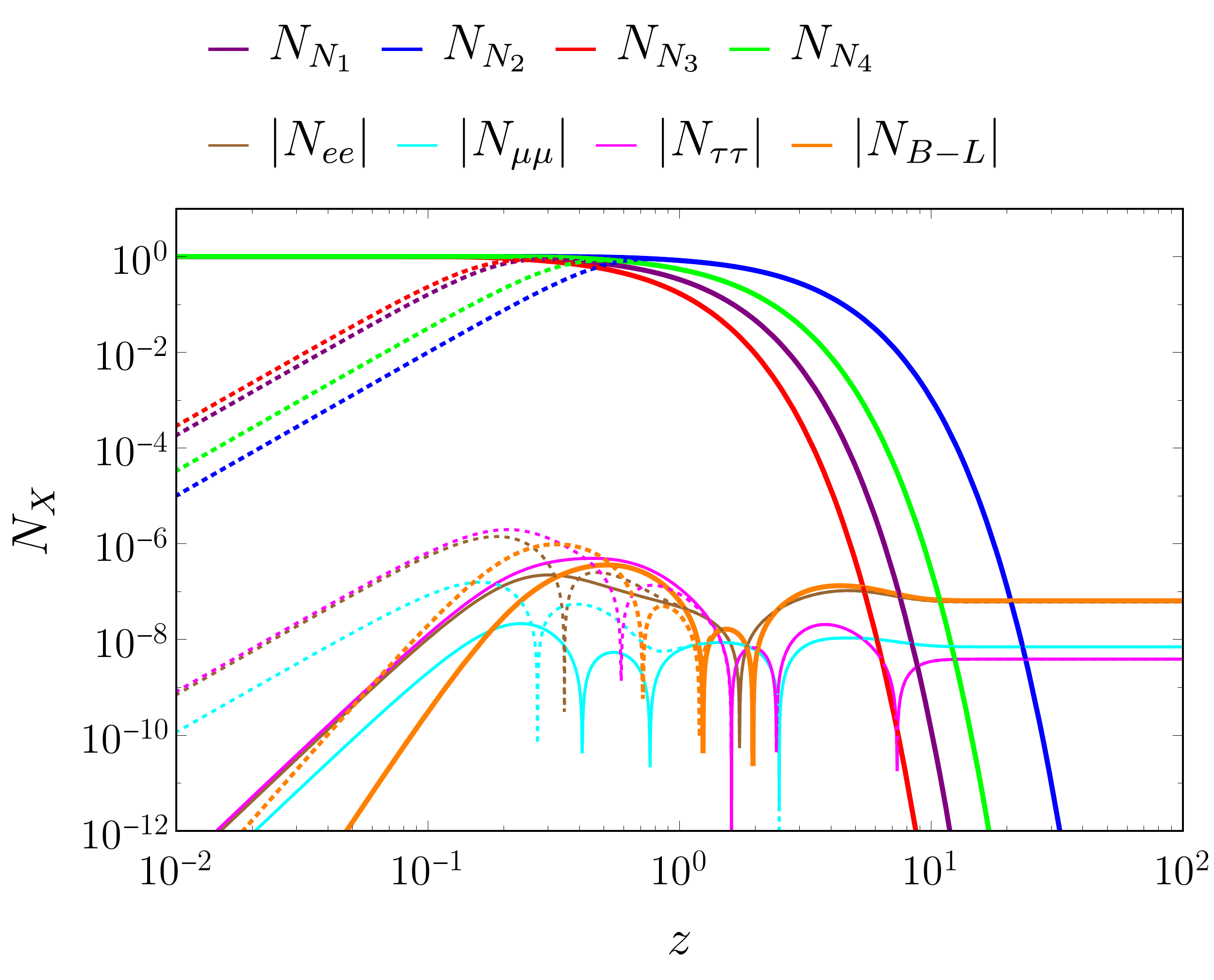}
    }\\
    \subfloat[$a = 2.4$ and $f = 2$]{
        \includegraphics[width=0.48\textwidth]{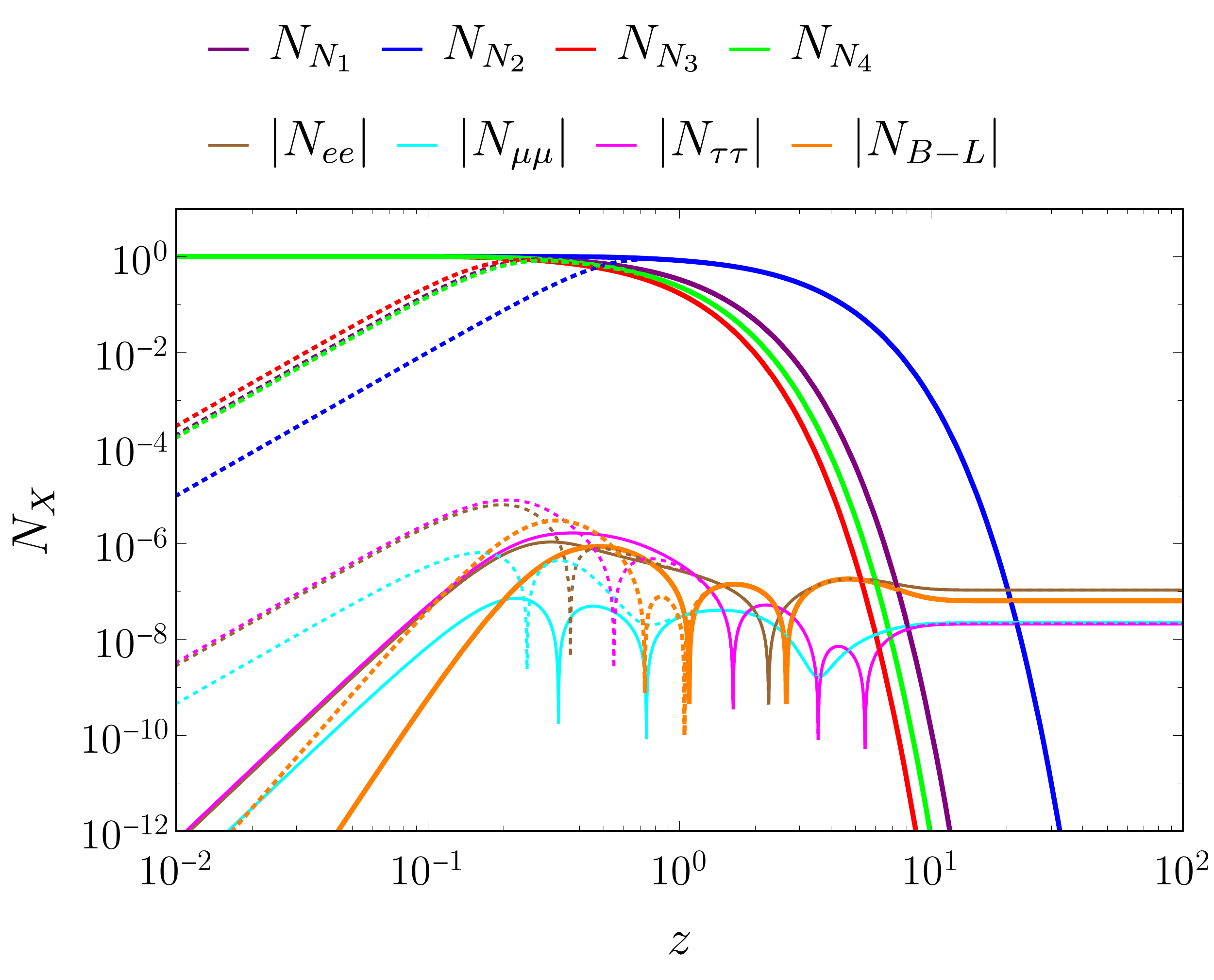}
    }
    \subfloat[$a = 3.3$ and $f = 2$]{
        \includegraphics[width=0.48\textwidth]{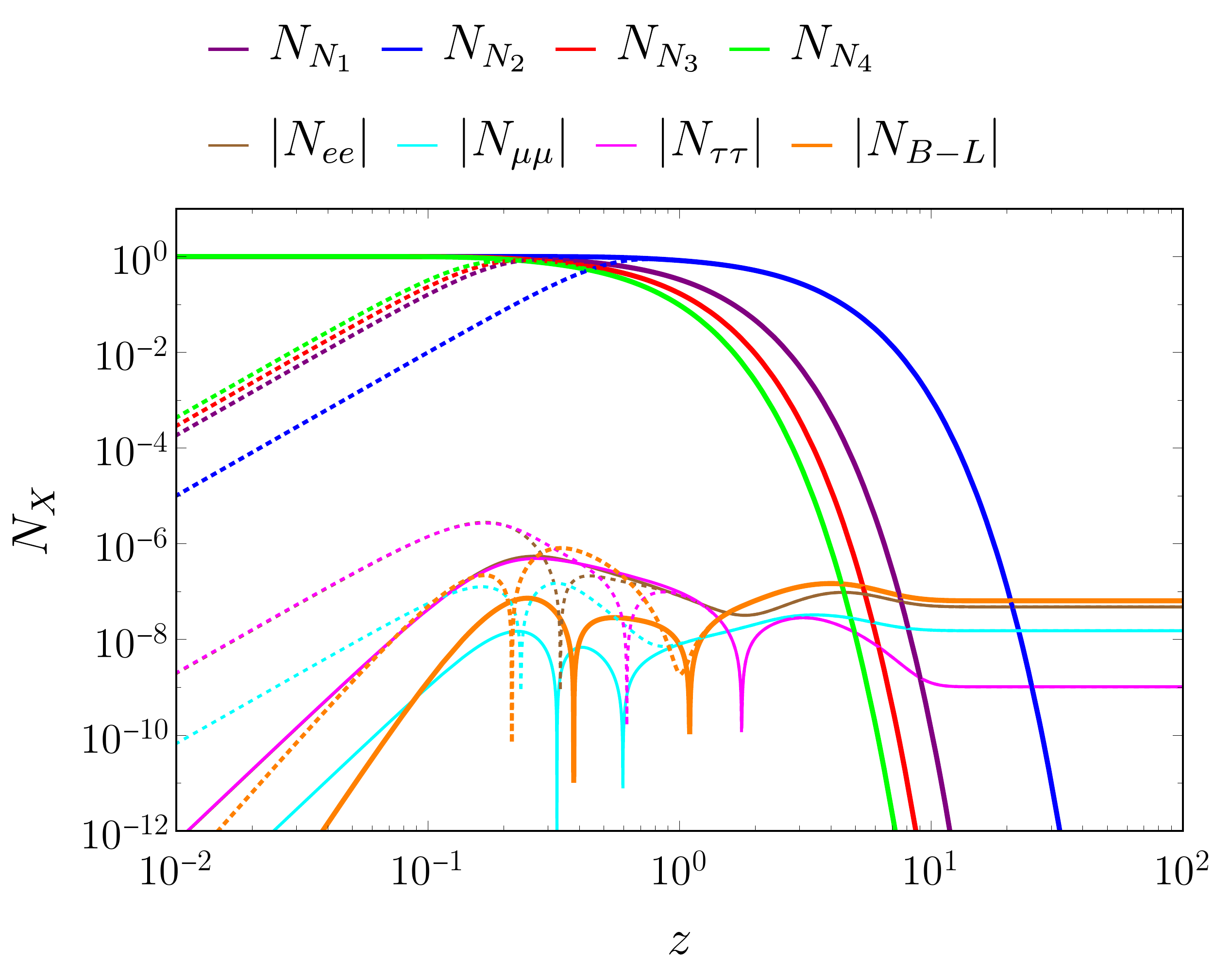}
    }
    \caption{Evolution of the Majorana neutrino number densities and the $B-L$ asymmetry for Case 1: $(b_1,b_2,b_3) \equiv (b(1,f,1)$. The dotted lines represent dynamical initial abundance $N_{N_i}(z=0) = 0$ and the solid lines represent thermal initial abundance $N_{N_i}(z=0) = N_{N_i}^{eq} (z=0)$. The final $B-L$ asymmetry does not depend on the initial conditions.}
    \label{NniNBL}
\end{figure}

We first solve equation \eqref{eq:Nni1} numerically to calculate the number densities $N_{N_i} (z)$, assuming both thermal initial abundance $N_{N_i}(z=0) = N_{N_i}^{eq} (z = 0)$, and dynamical initial abundance $N_{N_i}(z=0) = 0$. The results are shown in Figure \ref{NniNBL} for four representative cases: (a) $a = 0.3$ for region (i), (b) $a = 1.4$ for region (ii), (c) $a=2.4$ for region (iii), and (d) $a=3.3$ for region (iv). In both cases, the number densities at $z \gg 1$ are identical for both initial conditions. 
%

The number densities are then fed into the density matrix equation \eqref{eq:NBLdm} to calculate the $B-L$ asymmetry for the flavor components $N_{\alpha \beta}$ for the initial condition $N_{\alpha \beta}(z=0) = 0$. The trace of $N_{\alpha \beta}$ yields the total asymmetry. We find that the sign of the asymmetry is positive for $\delta = -78^\circ$ in all four regions. As shown in Figure \ref{NniNBL}, the final asymmetry does not depend on the initial conditions used to solve the first set of number density equations. 

The parameters of the density matrix equations are expressed in terms of the undetermined model parameters $a$ and $b$. For a particular value of $a$, we determine the value of $b$ that yields the $B-L$ asymmetry equal to the observed value, cf. Eqs.~ \eqref{measuredB} and \eqref{formulaB}:
\begin{align}
    N_{B-L}^f = 6.375 \times 10^{-8}.
\end{align}

\begin{figure}[!ht]
    \centering
    \subfloat[$f=2$ and $0.1 \leq a \leq 0.65$
    \label{fig:NBL2}]{
    \includegraphics[width=0.475\textwidth]{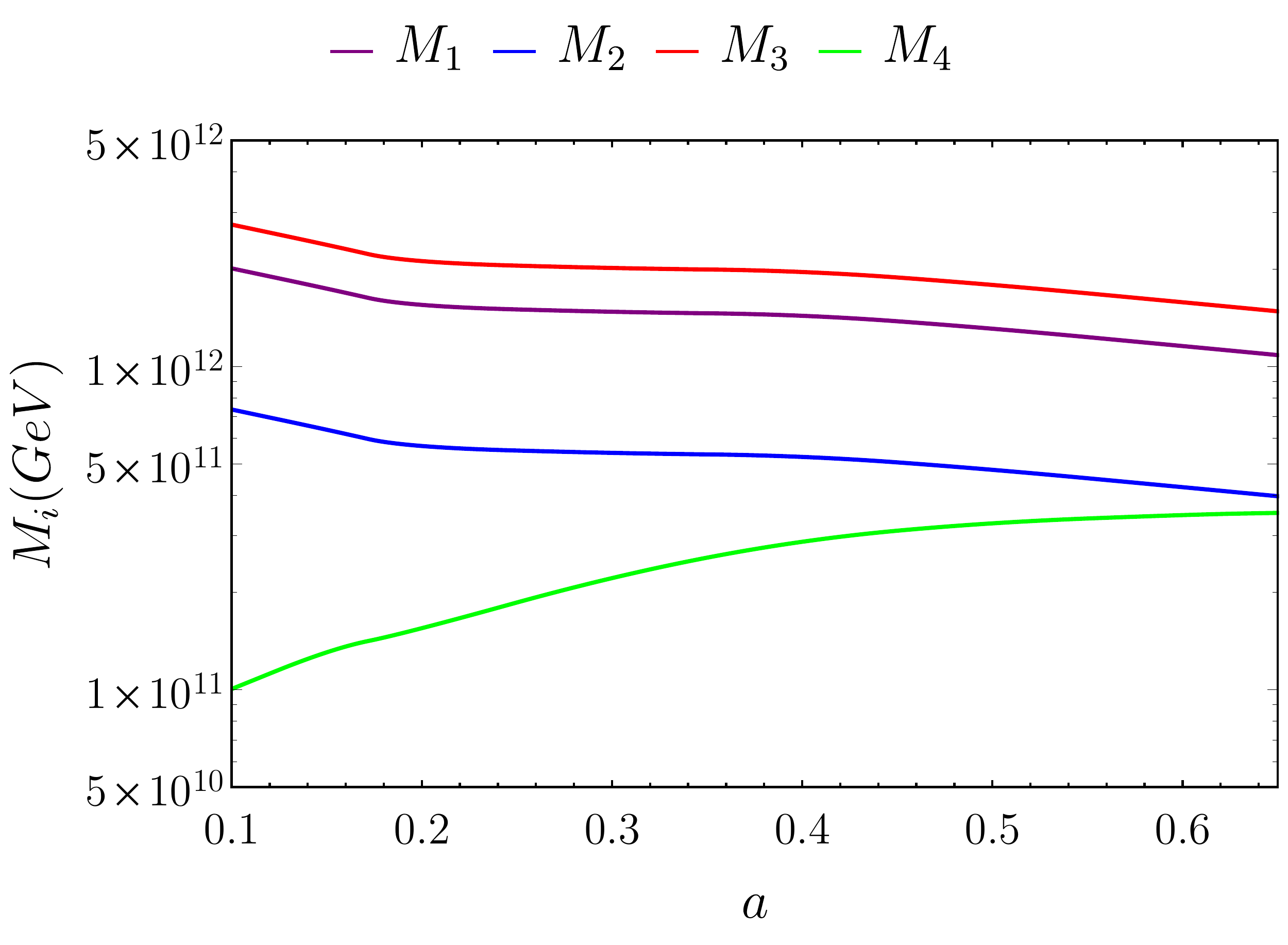}
    }
    \subfloat[$f=2$ and $0.8 \leq a \leq 1.9$
    \label{fig:Nmass2}]{
        \includegraphics[width=0.475\textwidth]{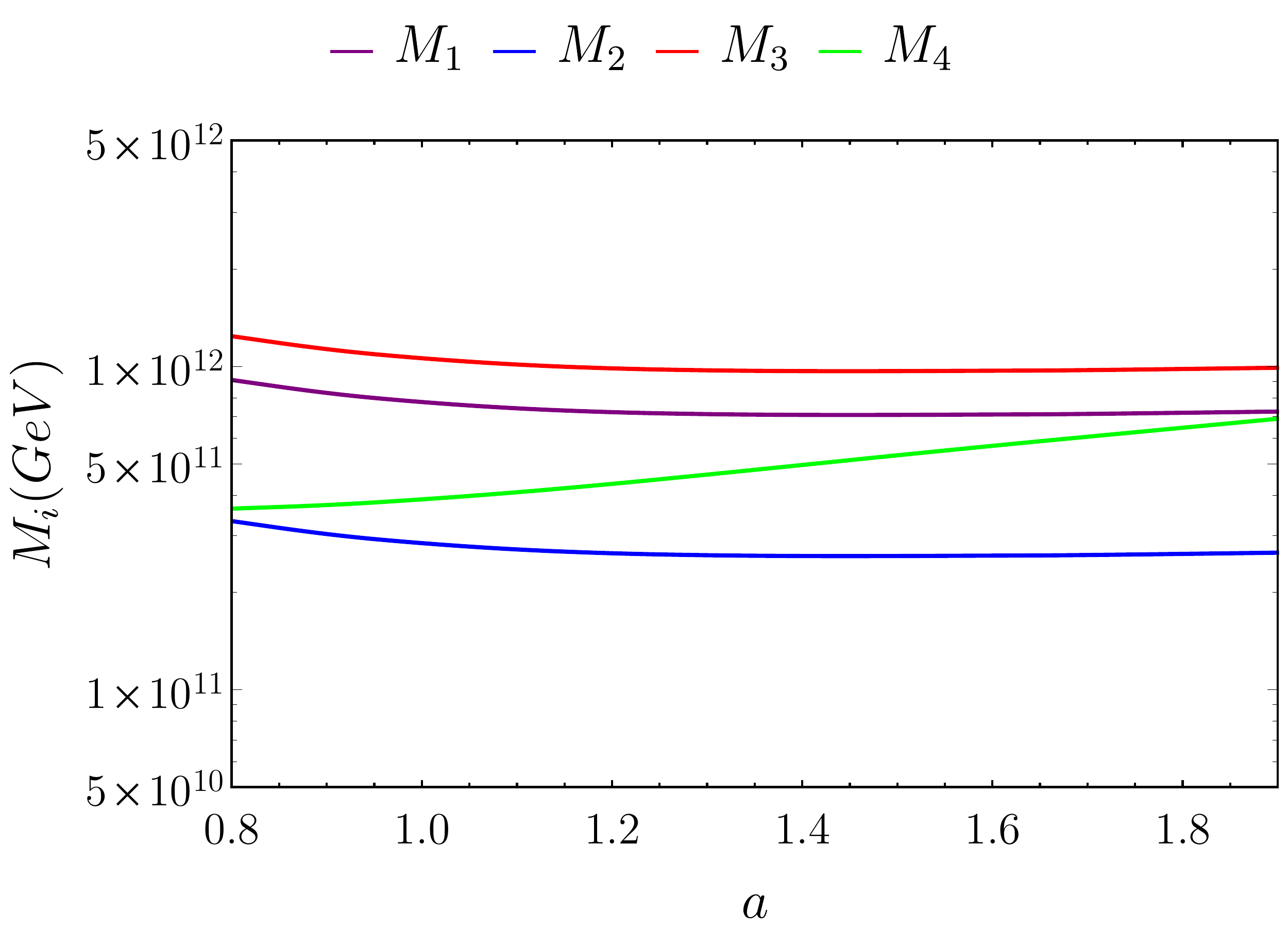}
    }\\
    \subfloat[$f=2$ and $2.1 \leq a \leq 2.65$
    \label{fig:NBL2}]{
    \includegraphics[width=0.475\textwidth]{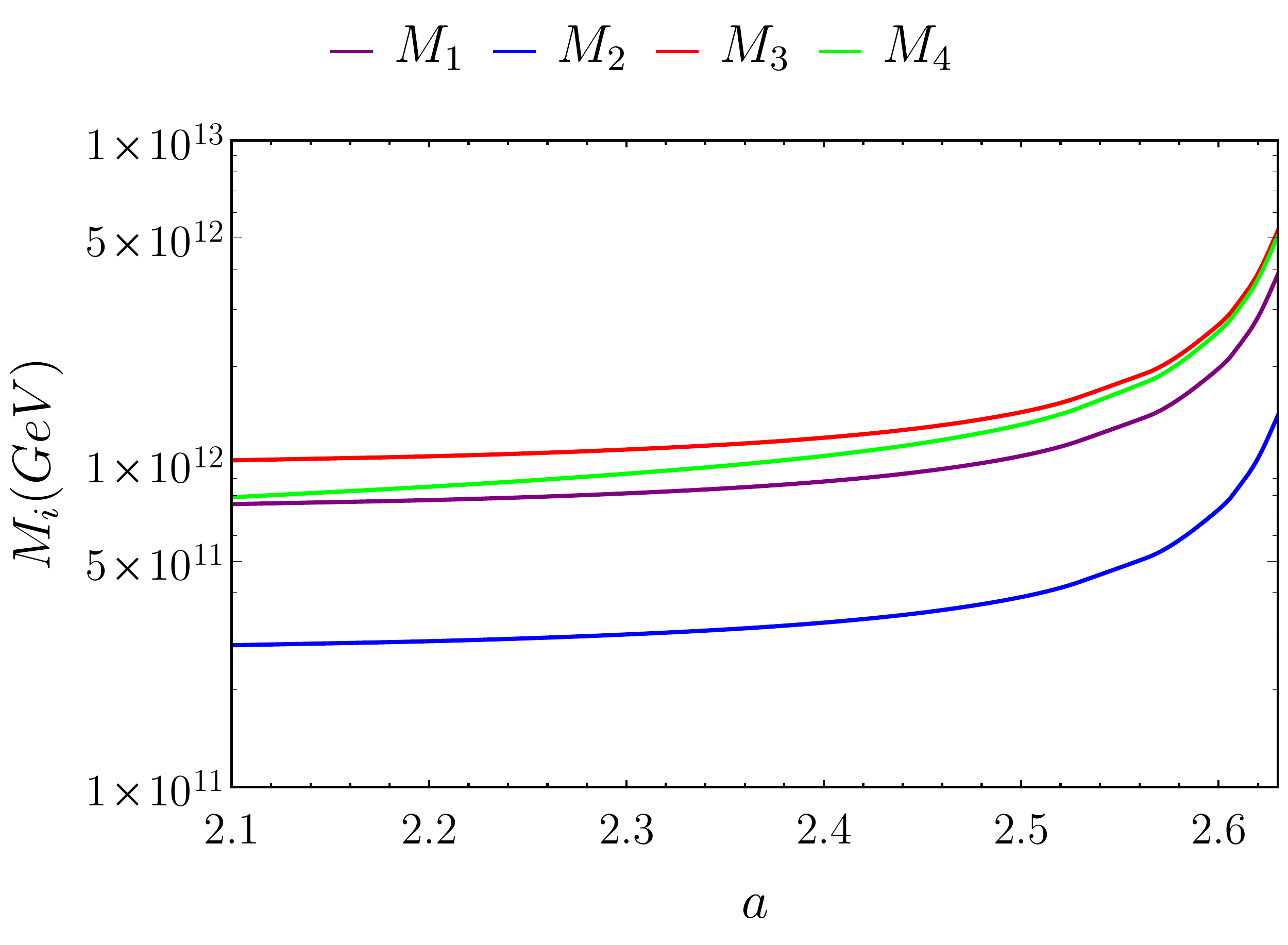}
    }
    \subfloat[$f=2$ and $2.85 \leq a \leq 5$ 
    \label{fig:Nmass2}]{
        \includegraphics[width=0.475\textwidth]{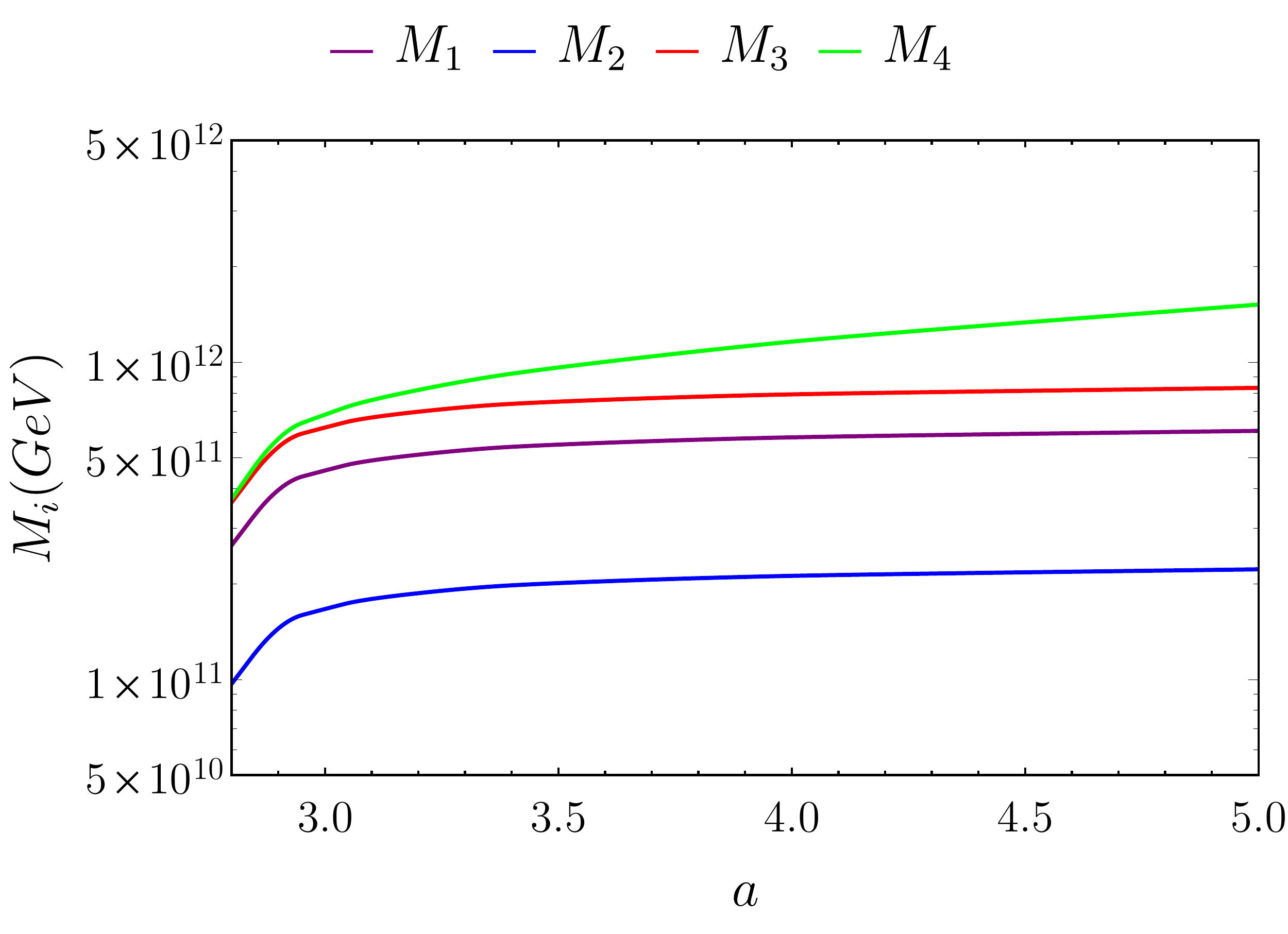}
    }
    \caption{
        Majorana mass spectrum for Case 1: $(b_1,b_2,b_3) \equiv (b(1,f,1)$ required to generate the observed baryon asymmetry for $f = 2$ in the regions (a) $0.1 \leq a \leq 0.65$, (b) $0.8 \leq a \leq 1.9$, (c) $2.1 \leq a \leq 2.65$ and (d) $2.85 \leq a \leq 5$.}
        \label{fig:NBLfin}
\end{figure}

Once $b$ is determined for a given $a$, we calculate the Majorana masses from Eq.~\eqref{Mmass_model1}. The mass spectrum for the four regions are shown in Figure \ref{fig:NBLfin}. For successful leptogenesis, the Majorana neutrino masses are of $\mc O(10^{11} - 10^{12})$ $\text{GeV} $. We notice that the masses could be much lower near the degeneracy between $M_3$ and $M_4$ at $a \simeq 2.73$ because of resonant effects. This is beyond the scope of the present work and will be explored in detail in a future paper.
%

\subsection{Case 2: $(b_1, b_2, b_3) \equiv b(f,f,1)$}
In this case the Dirac Yukawa matrix is given by:
\begin{align}
    Y^{(0)} &= \frac{\sqrt{b m_\nu}}{v} \left(
\begin{array}{cccc}
 0 & f & 0 & 2\beta\sqrt{f}   \\
 1 & 0 & 0 & -\beta\sqrt{f}  \\
 0 & 0 & - e^{i \delta} & \beta\sqrt{f} e^{i \delta}  \\
\end{array}
\right), \label{Y0new}
\end{align}
and the Majorana matrix is given by
\begin{align}
    \mc M &=  b\left(
\begin{array}{cccc}
 0 & f & 1 & 0 \\
 f & 0 & f & 0 \\
 1 & f & 0 & 0 \\
 0 & 0 & 0 & a \\
\end{array}
\right). \label{Majonew}
\end{align}
Its Takagi factorization yields
\begin{align}
\begin{split}
    M_1 &= b, \quad M_2 = \frac{b}{2} \left(\sqrt{8f^2+1}-1\right),\quad
    M_3 = \frac{b}{2} \left(\sqrt{8f^2+1}+1\right),\ \quad M_4 = a b, \label{Mmass}
\end{split}
\end{align}
and
\begin{align}
    \mc U_m = \left(
\begin{array}{cccc}
 -\frac{i}{\sqrt{2}} & \frac{-i}{2} \sqrt{1-\frac{1}{\sqrt{8f^2+1}}} & \frac{1}{2} \sqrt{1+\frac{1}{\sqrt{8f^2+1}}} & 0 \\ [1.5em]
 \frac{i}{\sqrt{2}} & \frac{-i}{\sqrt{2}} \sqrt{1-\frac{1}{\sqrt{8f^2+1}}} & \frac{1}{\sqrt{2}} \sqrt{1+\frac{1}{\sqrt{8f^2+1}}} & 0 \\[1em]
 0 & \frac{i}{2}  \sqrt{1+\frac{1}{\sqrt{8f^2+1}}} & \frac{1}{2} \sqrt{1-\frac{1}{\sqrt{8f^2+1}}} & 0 \\[1.5em]
 0 & 0 & 0 & 1 \\
\end{array}
\right). \label{Unitmnew}
\end{align}

\begin{figure}[!ht]
    \centering
    \includegraphics[width=0.48\textwidth]{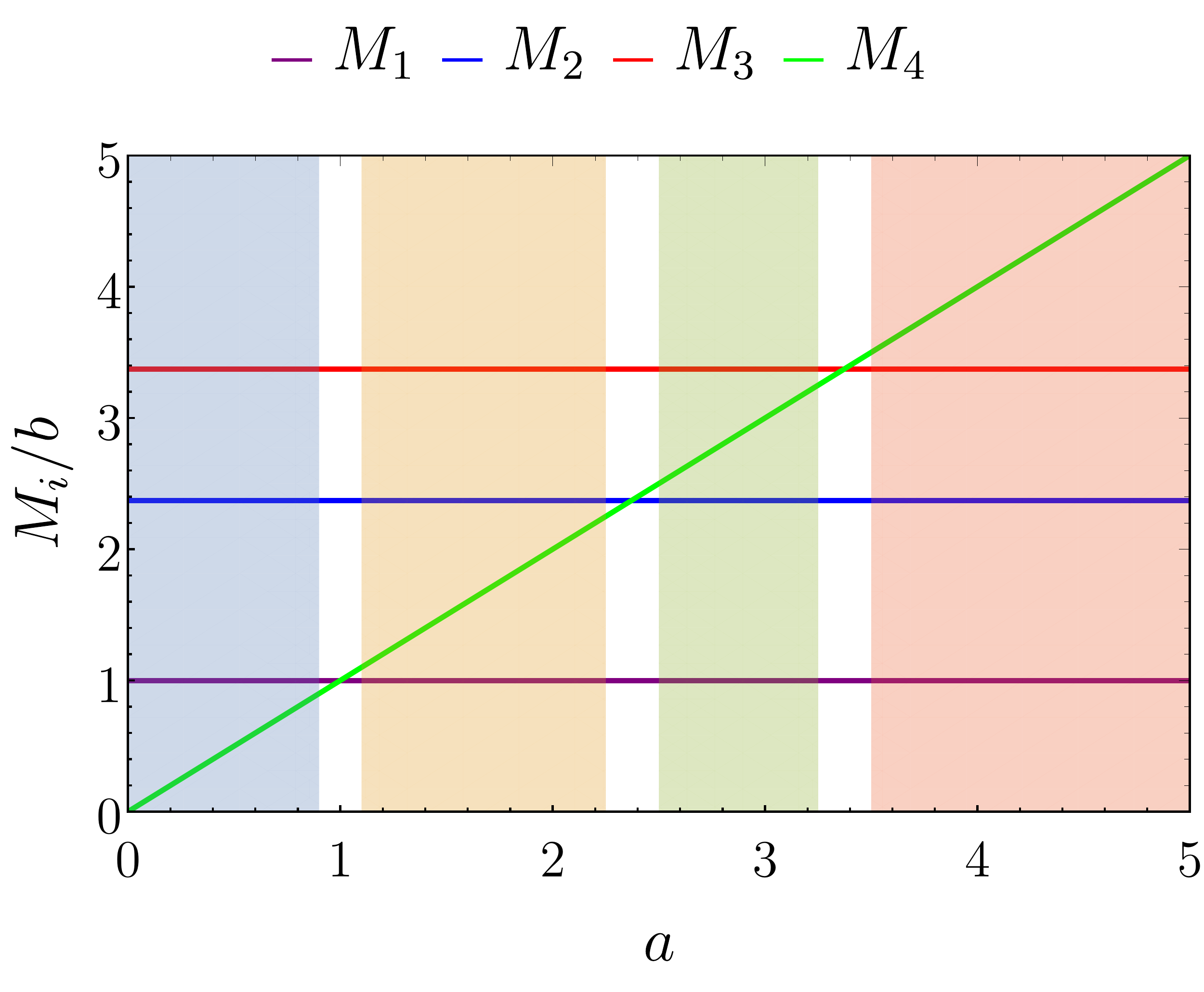}
\caption{Majorana neutrino mass spectrum for Case 2: $(b_1,b_2,b_3) \equiv (b(f,f,1)$. $M_1$, $M_2$ and $M_3$ do not depend on $a$. $M_4$ is degenerate with $M_1$, $M_2$ and $M_3$ for $a \simeq 1, 2.37$ and $3.37$, respectively, setting $f=2$. The parameter space can be divided into four regions to avoid near-degeneracies. } 
\label{fig:massnew2}
\end{figure}

As before we will set $f=2$ for the remainder of our discussion whenever a numerical value is required, although we have verified that the results are equivalent when $f \sim \mc O(1)$ and $f\neq 1$. The Majorana mass spectrum is given by:
\begin{align}
    \frac{M_1}{b} = 1,\quad \frac{M_2}{b} = \frac{1}{2}(\sqrt{33}-1), \quad \frac{M_3}{b} = \frac{1}{2}(\sqrt{33}+1), \quad \frac{M_4}{b} = a. \label{Mmass2}
\end{align}
$M_4$ can be degenerate with $M_1$, $M_2$ and $M_3$ for $a = 1$, $\frac{1}{2}(\sqrt{33}-1) \simeq 2.37$ and $\frac{1}{2}(\sqrt{33}+1) \simeq 3.37$, respectively, as shown in Figure \ref{fig:massnew2}. To avoid resonant enhancement near degeneracies we split the parameter space into four regions: (i) $0.1 \leq a \leq 0.9$, (ii) $1.1 \leq a \leq 2.25$, (iii) $2.5 \leq a \leq 3.25$, and (iv) $a \geq 3.5$, shown in Figure \ref{fig:massnew2}.  
%

\begin{figure} 
    \centering
    \subfloat[$a = 0.35$ and $f = 2$]{
        \includegraphics[width=0.48\textwidth]{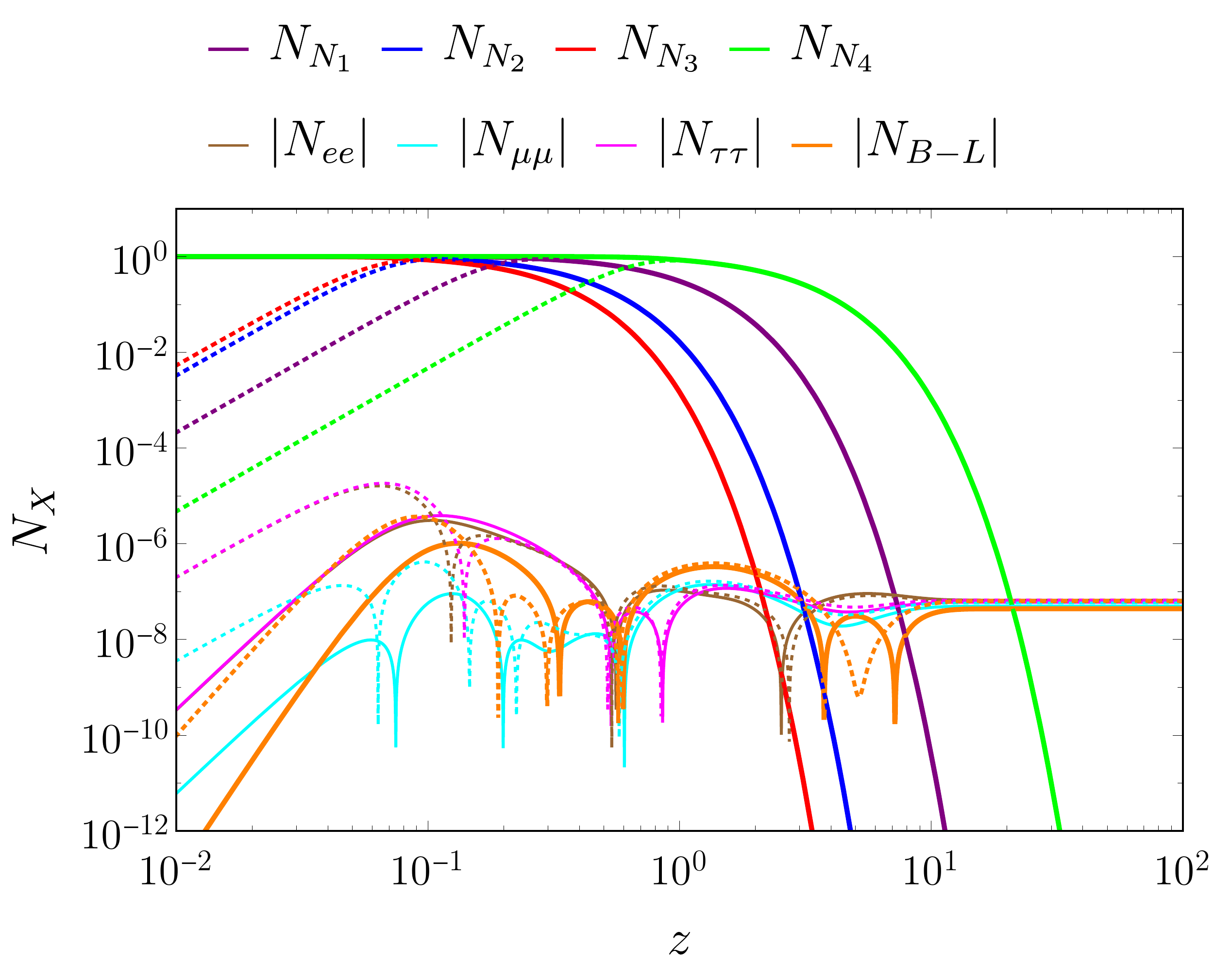}
    }
    \subfloat[$a = 0.36$ and $f = 2$]{
        \includegraphics[width=0.48\textwidth]{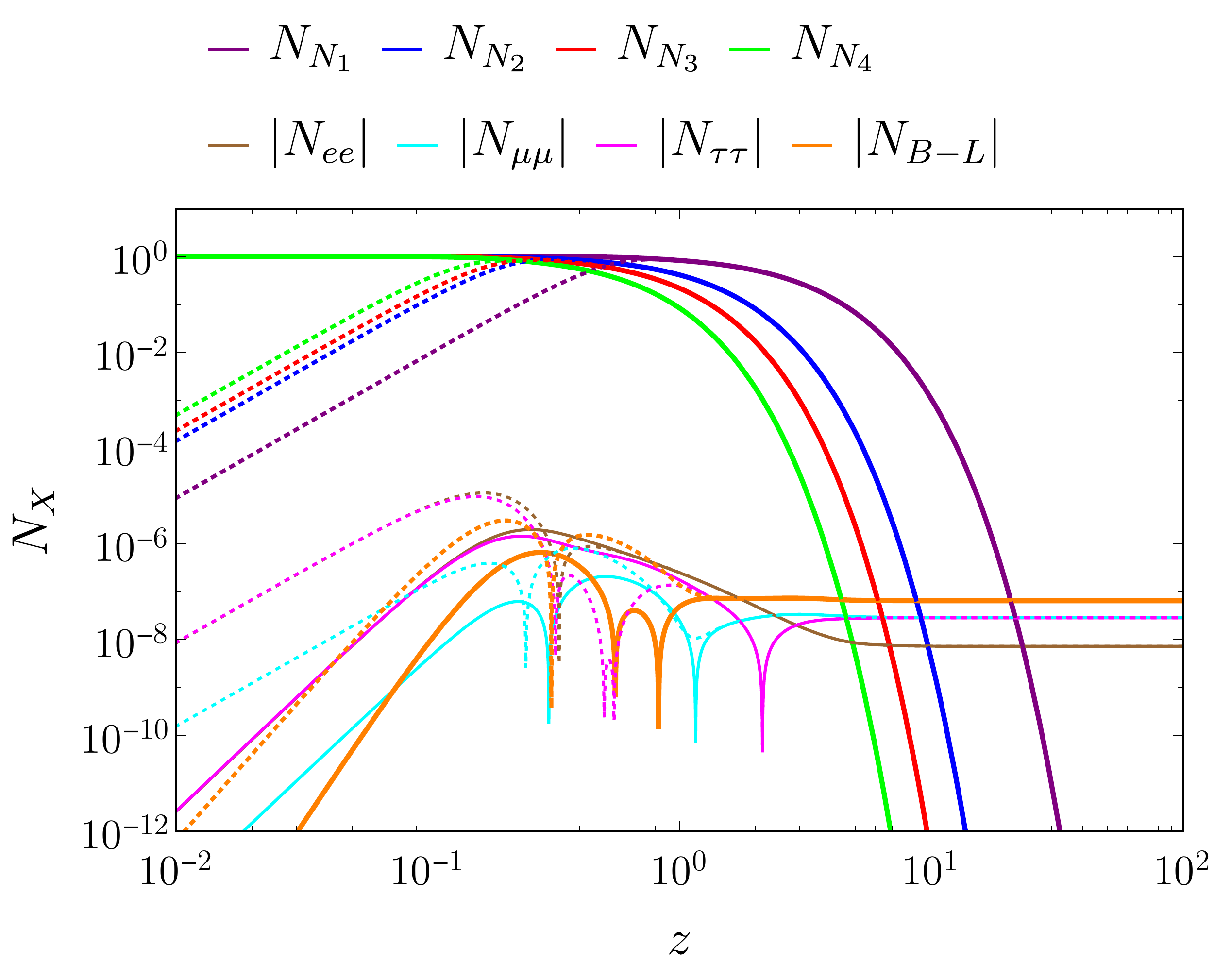}
    }\\
    \subfloat[$a = 1.7$ and $f = 2$]{
        \includegraphics[width=0.48\textwidth]{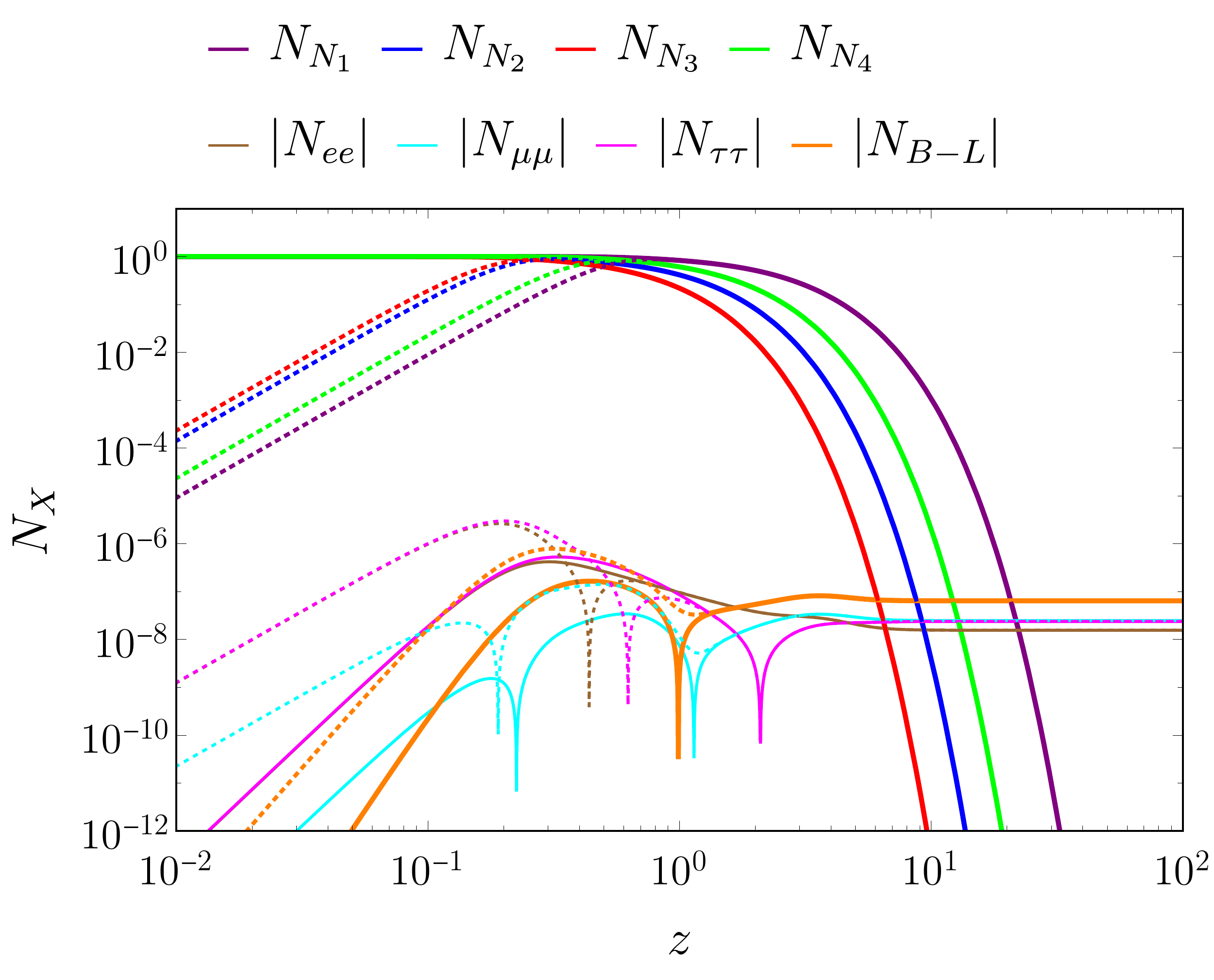}
    }
    \subfloat[$a = 3.25$ and $f = 2$]{
        \includegraphics[width=0.48\textwidth]{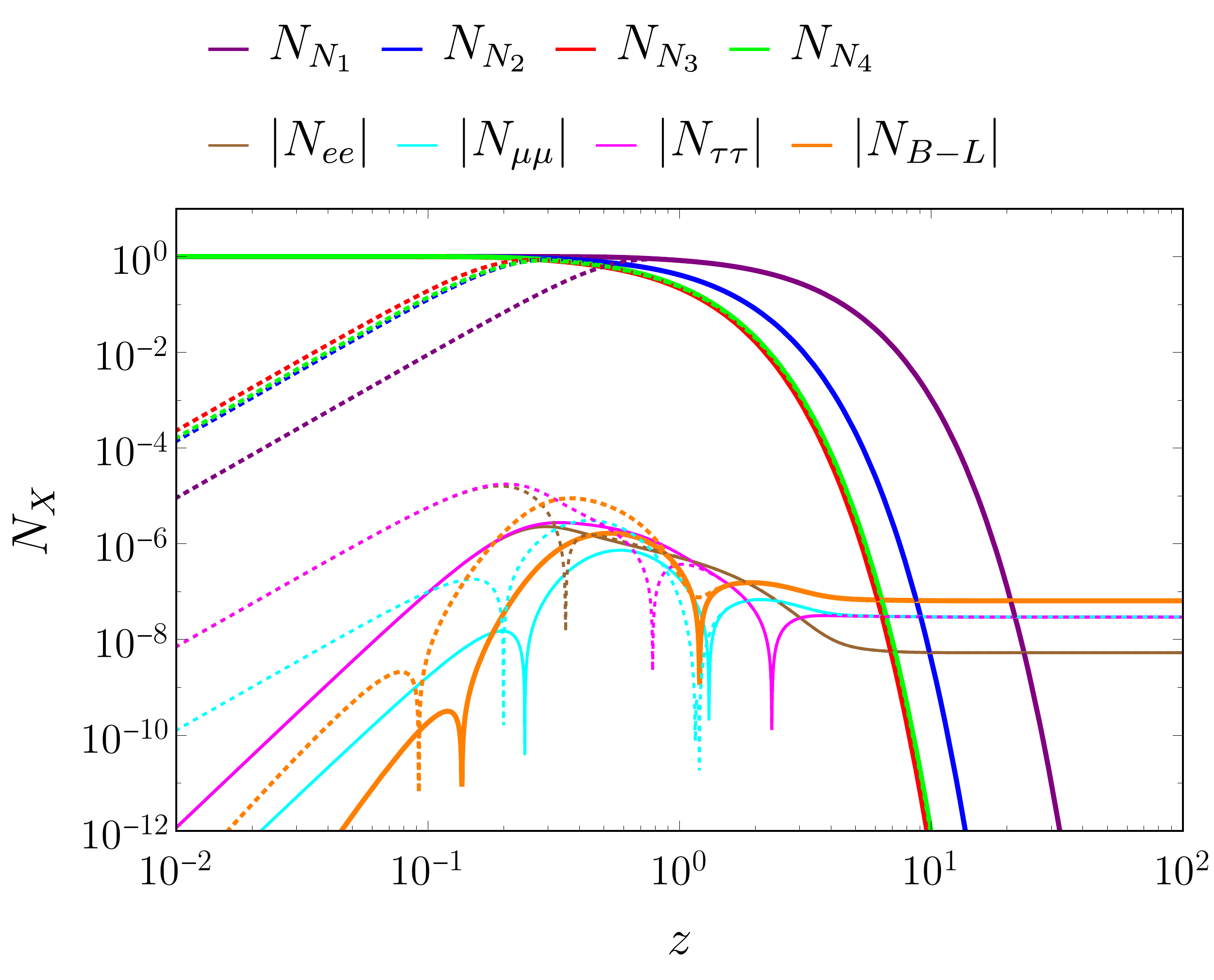}
    }\\
    \subfloat[$a = 3.5$ and $f = 2$]{
        \includegraphics[width=0.48\textwidth]{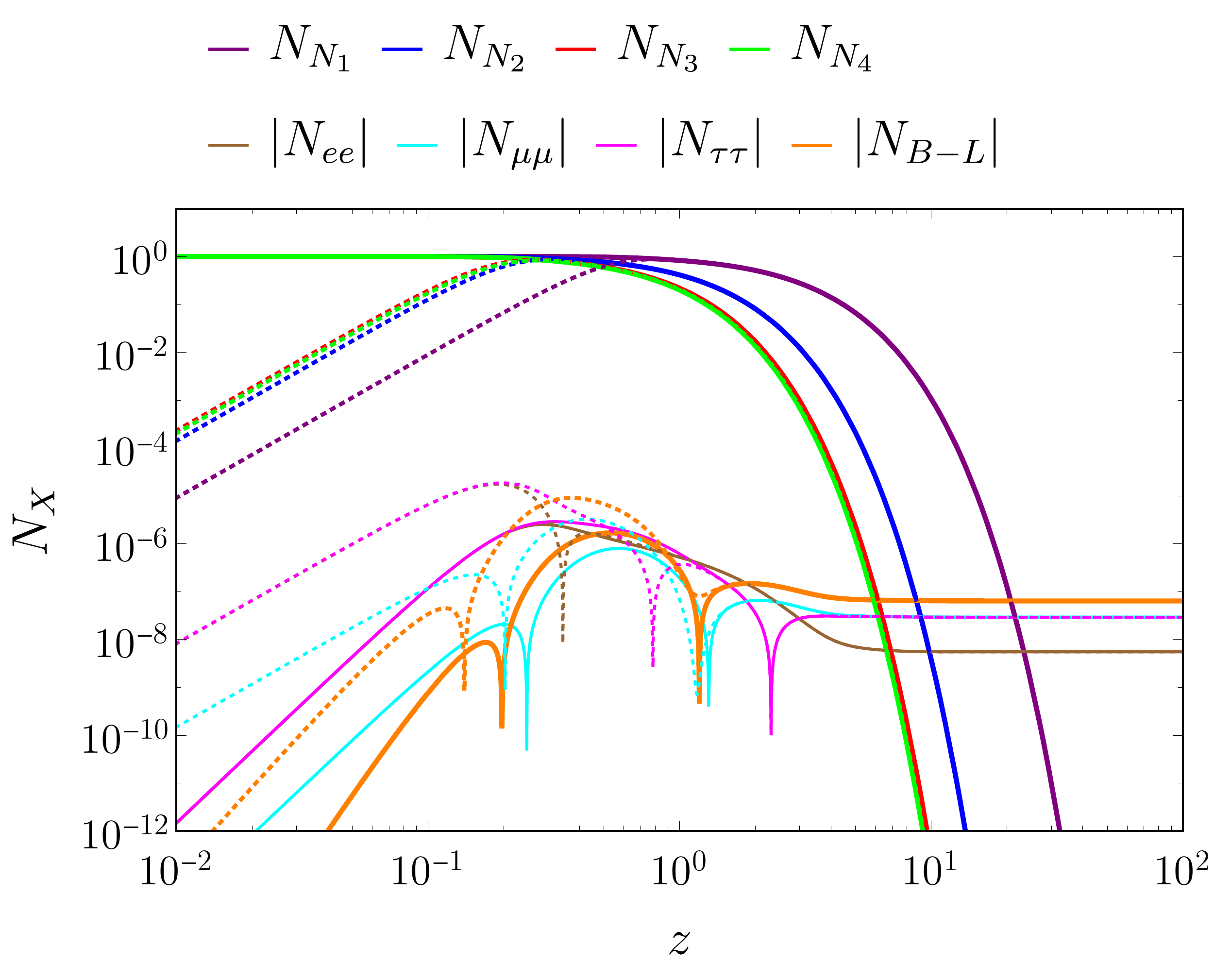}
    }
    \caption{Evolution of the Majorana neutrino number densities and the $B-L$ asymmetry for Case 2: $(b_1,b_2,b_3) \equiv (b(f,f,1)$. The dotted lines represent dynamical initial abundance $N_{N_i}(z=0) = 0$ and the solid lines represent thermal initial abundance $N_{N_i}(z=0) = N_{N_i}^{eq} (z=0)$. The final $B-L$ asymmetry does not depend on the initial conditions.}
    \label{NBL_model2}
\end{figure}

The decay parameters in this case are $K_1 \simeq 54.02$, $K_2 \simeq 62.87$, $K_3 \simeq 35.86$ and $K_4 \simeq 28.21$, implying strong washout and hence justifying our use of the density matrix formalism. As before, we numerically solve the density matrix equations to determine $b$ for a particular value of $a$ so that the generated $B-L$ asymmetry is equal to the observed value from CMB. We find that the $B-L$ asymmetry is positive for $\delta=-78^\circ$ at $0.1 \leq a \leq 0.35$ and $3.5 \leq a \leq 5$ and for $\delta = 78^\circ$ at $0.36 \leq a \leq 0.9$, $1.1 \leq a \leq 2.25$ and $2.5 \leq a \leq 3.25$. The number densities and $B-L$ asymmetries for representative values of $a$ are shown in Figure \ref{NBL_model2}.

\begin{figure}
    \centering
    \subfloat[$f=2$ and $0.1 \leq a \leq 0.35$
    \label{fig:NBL2}]{
    \includegraphics[width=0.475\textwidth]{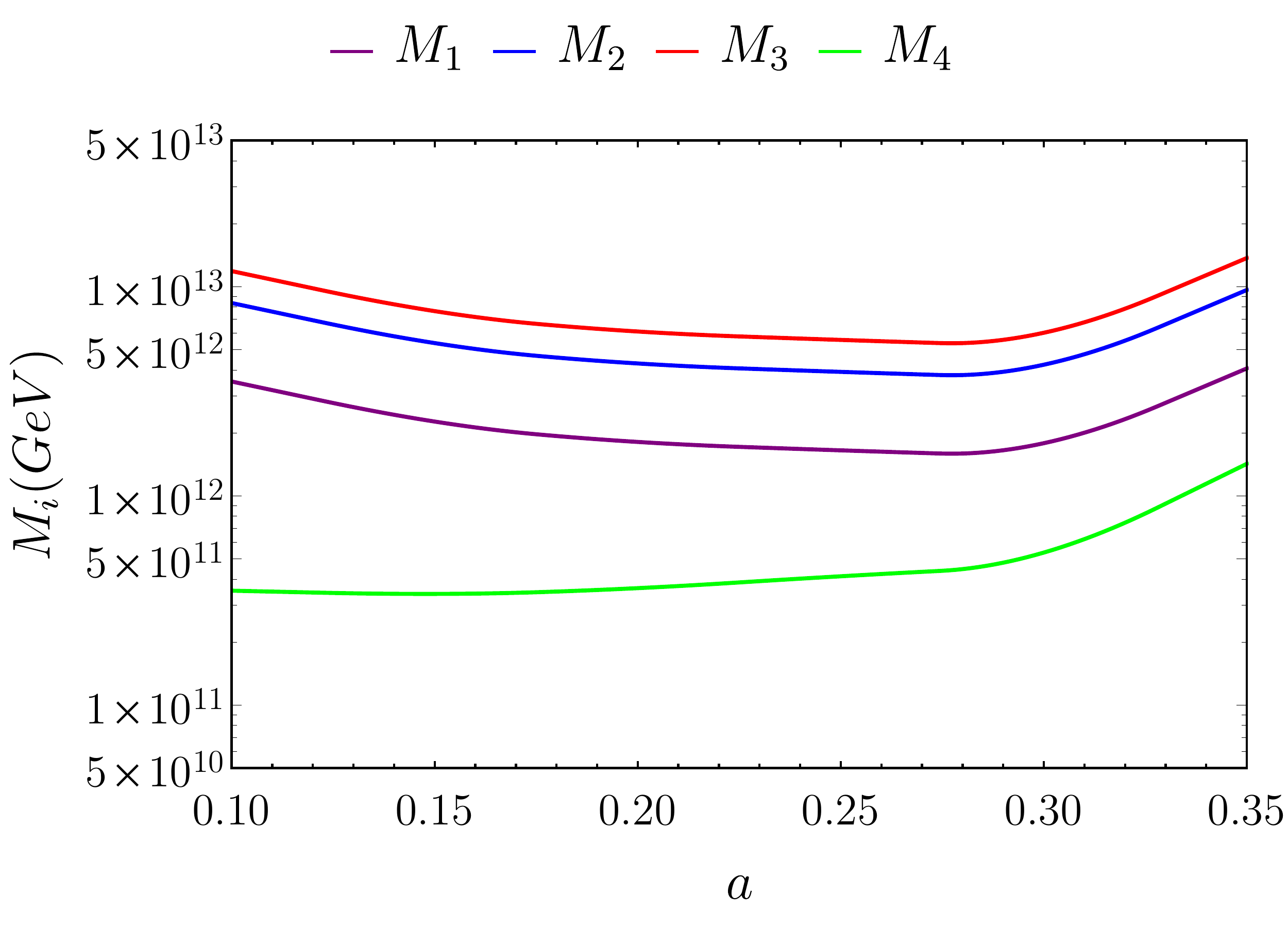}
    }
    \subfloat[$f=2$ and $0.36 \leq a \leq 0.9$
    \label{fig:NBL2}]{
    \includegraphics[width=0.475\textwidth]{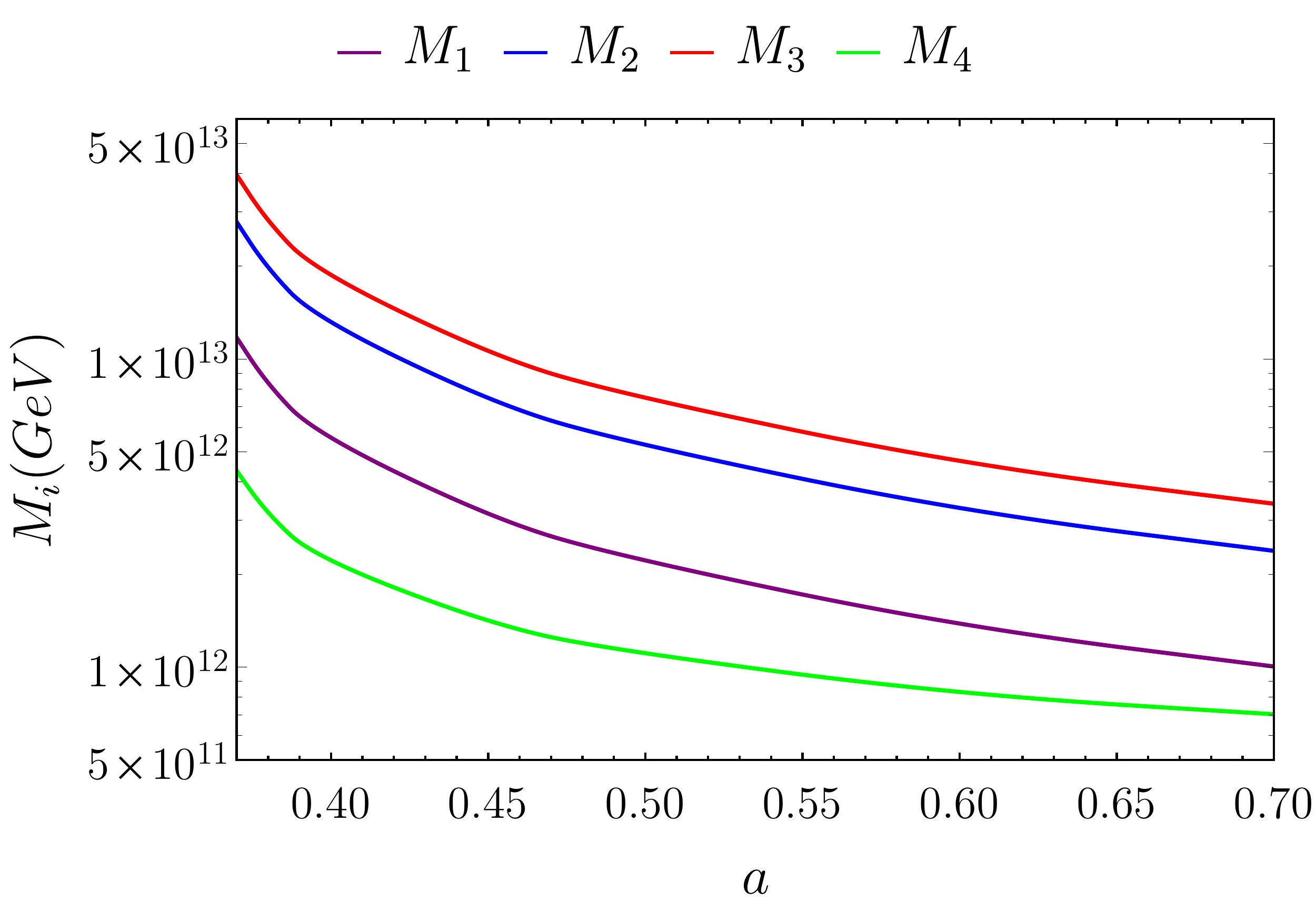}
    }\\
    \subfloat[$f=2$ and $1.1 \leq a \leq 2.25$
    \label{fig:Nmass2}]{
        \includegraphics[width=0.475\textwidth]{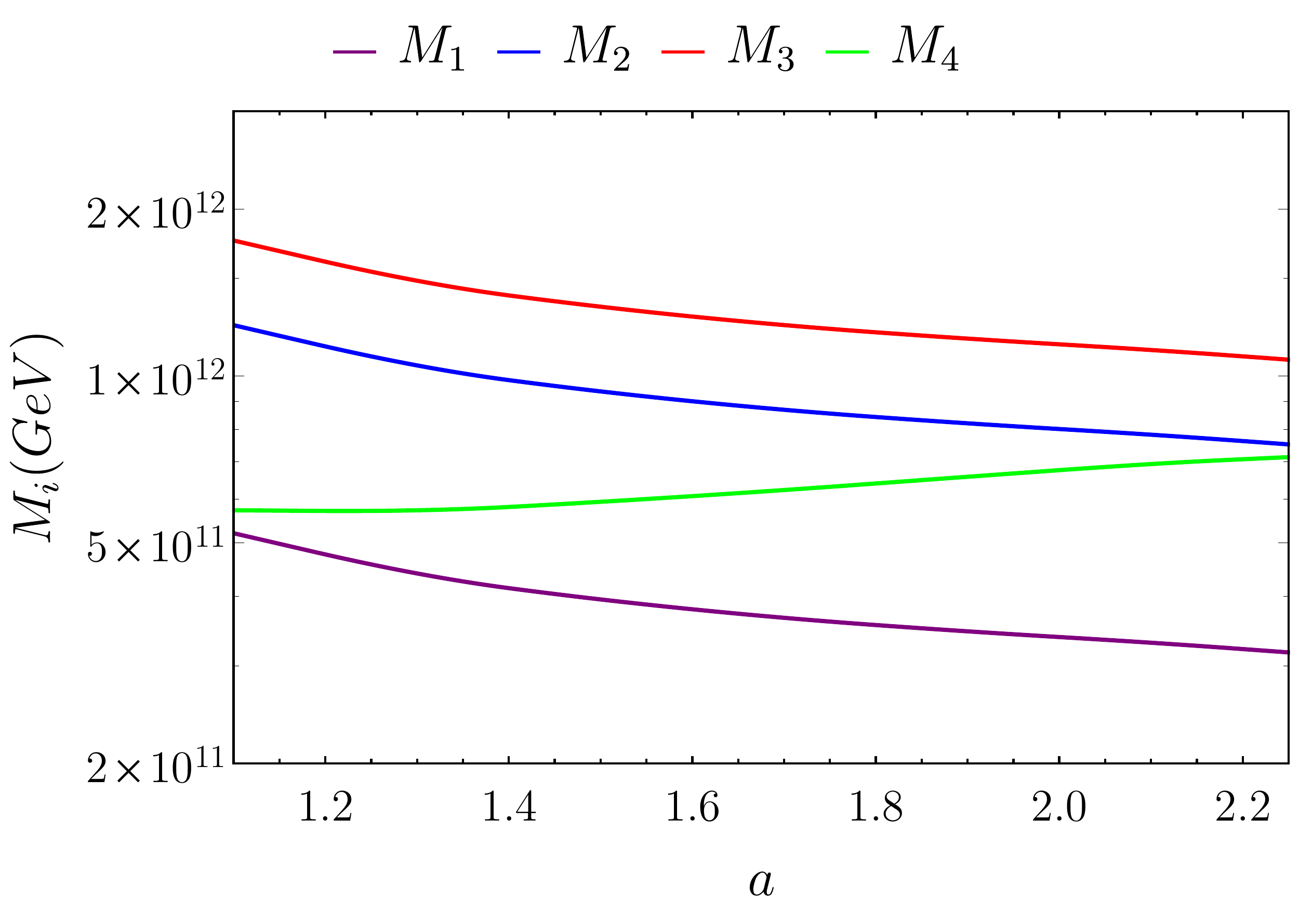}
    }
    \subfloat[$f=2$ and $2.5 \leq a \leq 3.25$
    \label{fig:NBL2}]{
    \includegraphics[width=0.475\textwidth]{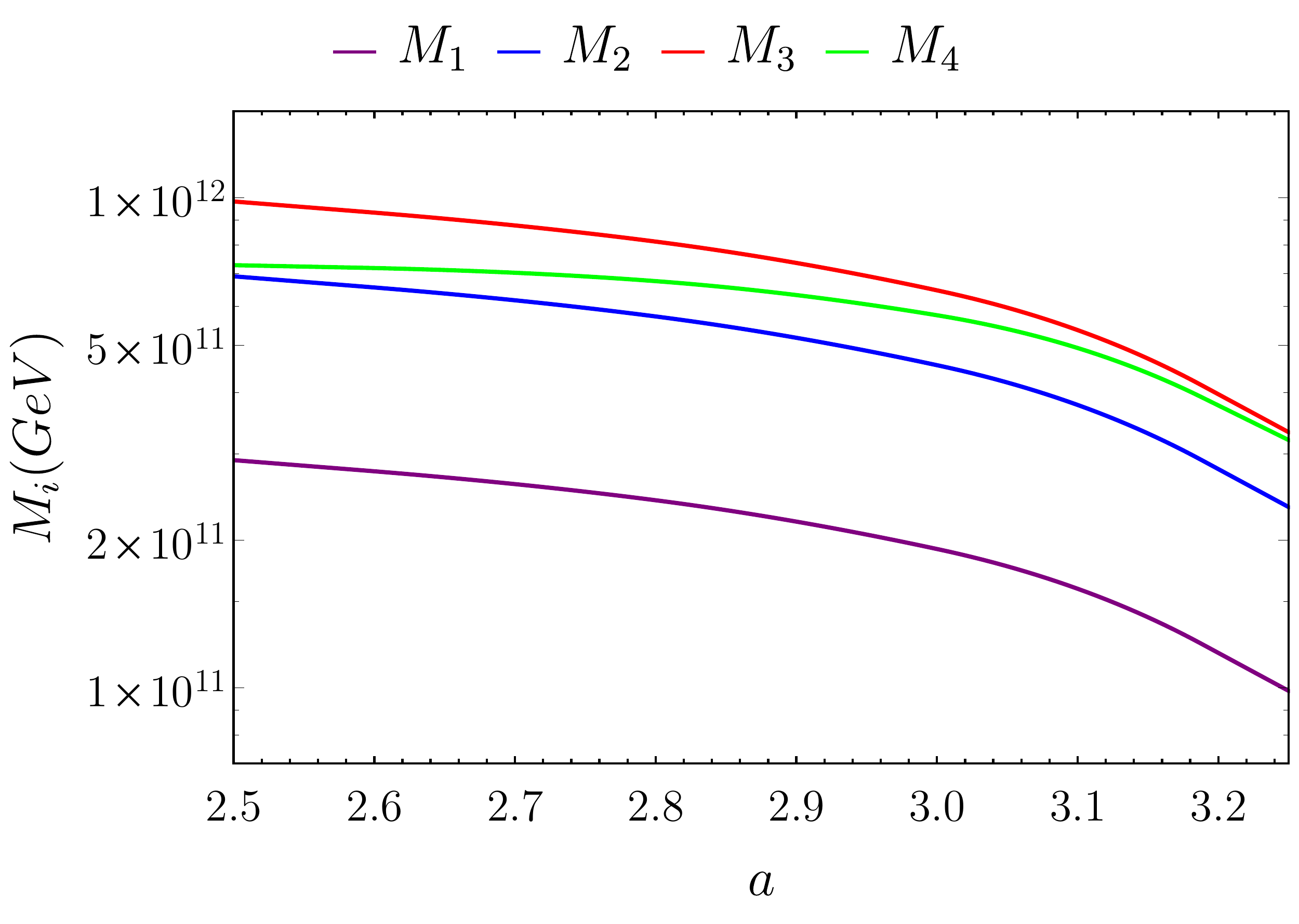}
    }\\
    \subfloat[$f=2$ and $3.5 \leq a \leq 5$ 
    \label{fig:Nmass2}]{
        \includegraphics[width=0.475\textwidth]{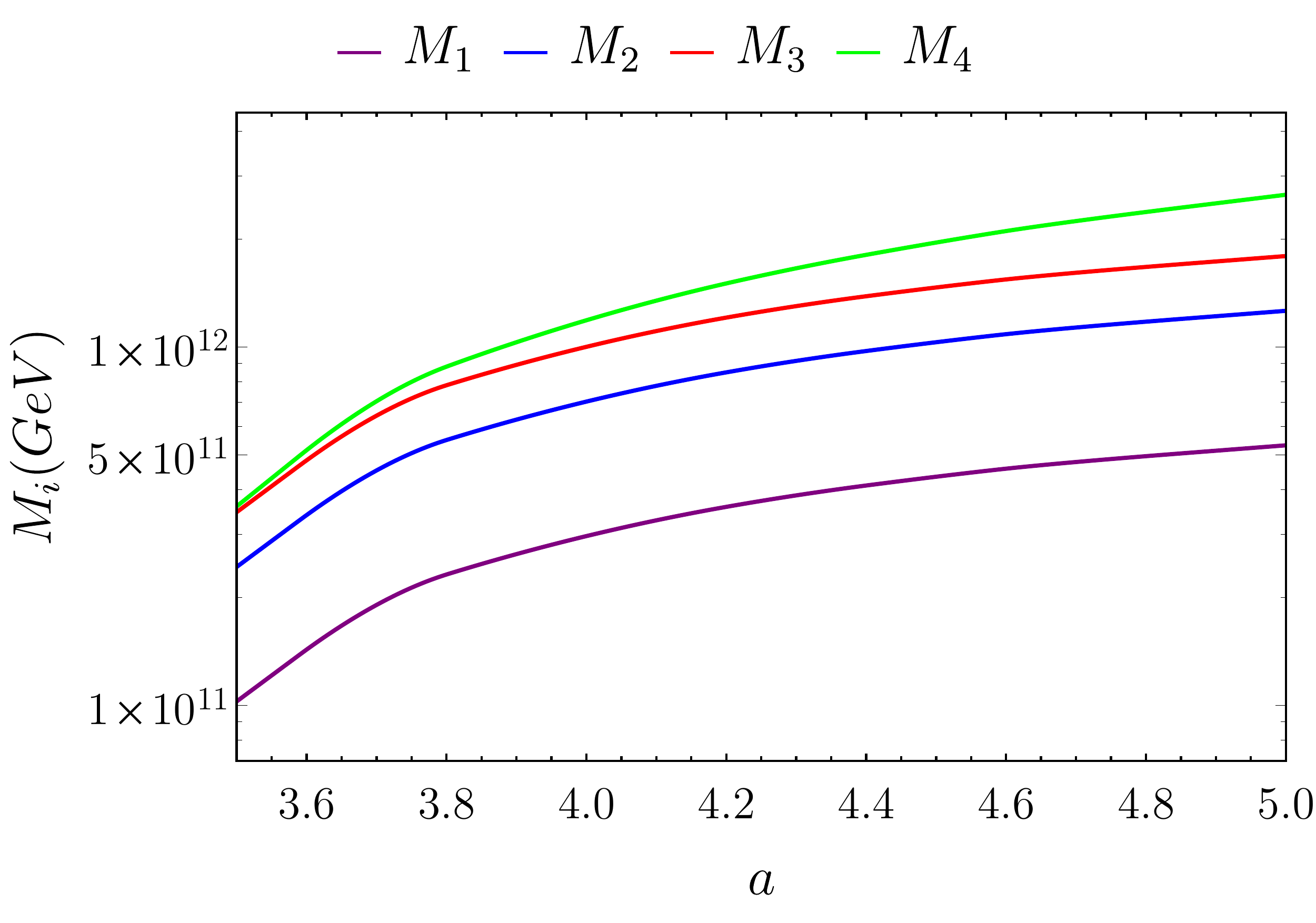}
    }
    \caption{
        Majorana mass spectrum for Case 2: $(b_1,b_2,b_3) \equiv (b(f,f,1)$ required to generate the observed baryon asymmetry for $f = 2$ in the regions (a) $0.1 \leq a \leq 0.35$, (b) $0.36 \leq a \leq 0.9$, (c) $1.1 \leq a \leq 2.25$, (d) $2.5 \leq a \leq 3.25$ and (e) $3.5 \leq a \leq 5$.}
        \label{Mmass_model2}
\end{figure}

The Majorana mass spectrum required to generate the observed baryon asymmetry is shown in Figure \ref{Mmass_model2}. The masses are of $\mc O(10^{11} - 10^{12})\ \text{GeV}$ for most part of the parameter space. We notice that the masses seem to be decreasing near the degeneracy between $M_3$ and $M_4$ at $a \simeq 3.37$. Since resonant effects are beyond the scope of this work, we leave the investigation of this possibility for a future work.
%

\section{Sign of the $\cancel{CP}$ phases and the baryon asymmetry} \label{sec:5}
In the asymmetric texture, the TBM phase is determined as ${\delta} \simeq \pm 78^\circ$ \cite{Rahat:2018sgs} from the requirement to match the observed reactor angle.\footnote{See appendix \ref{app:robust} for a discussion on the robustness of the leptogenesis results when $\delta$ is allowed to vary in the range that still reproduces all three PMNS angles within $3\sigma$ of their PDG value.} The sign ambiguity in $\delta$ is not resolved from the physics of the electroweak sector. In this section we will explore the possibility of relating this sign to the sign of the baryon asymmetry. 

The dependence on $\delta$ comes through the neutrino Yukawa matrix $Y_\nu$, which appears in the decay parameter $K_{i}$ and the $CP$-asymmetry parameter $\varepsilon_{\alpha \beta}^{(i)}$. From Eq.~\eqref{YYCI}, it goes away in  $(Y_\nu^\dagger Y_\nu)_{ii}$; hence $K_{i}$ in Eq.~\eqref{Ki} does not depend on $\delta$. This implies that the number density of the Majorana neutrinos in Eq.~\eqref{eq:Nni1} can be determined independently of $\delta$.

The $B-L$ asymmetry is the sum of the diagonal terms $N_{\alpha \alpha}$, which are proportional to $CP$ asymmetries $\varepsilon_{\alpha \alpha}^{(i)}$. From Eq.~\eqref{epsidm}, this depends on $\text{Im}\left[ (Y_\nu^*)_{\alpha i} (Y_\nu)_{\alpha j} (Y_\nu^\dagger Y_\nu)_{ij} \right]$ and $\text{Im}\left[ (Y_\nu^*)_{\alpha i} (Y_\nu)_{\alpha j} (Y_\nu^\dagger Y_\nu)_{ji} \right]$. We write $Y_\nu$ as 
\begin{align}
    Y_\nu &= \mc U^{(-1)\dagger} \text{diag}(1,1,e^{i\delta}) W^T P \mc D_m^{1/2} \label{Ynuform}
\end{align}
following Eqs.~\eqref{Rform} and \eqref{YnuCI}. 
Then Eqs.~\eqref{Ynuform} and \eqref{YYCI} yield
\begin{align}
    \text{Im}\left[ (Y_\nu^*)_{\alpha i} (Y_\nu)_{\alpha j} (Y_\nu^\dagger Y_\nu)_{ij} \right] &=  {P^*}_{ii}^2 {P^*}_{jj}^2  (\mc D_m)_{ii} (\mc D_m)_{jj}
    \sum_{\beta, \gamma, \kappa} \mc U^{(-1)}_{\beta\alpha} \mc U^{(-1)}_{\gamma\alpha} W_{i\beta} W_{j\gamma} W_{i\kappa} W_{j\kappa} \nonumber \\
    & \times \text{Im} \left[ (\text{diag}(1,1,e^{-i\delta}))_{\beta \beta} (\text{diag}(1,1,e^{i\delta}))_{\gamma \gamma} \right],
\end{align}
where $j \neq i$. The imaginary part on the right-hand side is nonzero, and proportional to $\sin \delta$, when either $\beta =3$ or $\gamma=3$. Similar arguments apply for $\text{Im}\left[ (Y_\nu^*)_{\alpha i} (Y_\nu)_{\alpha j} (Y_\nu^\dagger Y_\nu)_{ji} \right]$. Therefore the diagonal $CP$-asymmetry parameters and the final $B-L$ asymmetry are proportional to  $\sin \delta$, and demanding that the calculated asymmetry has a positive sign fixes the sign of $\delta$. 


For $\delta \simeq -78^\circ$, the Dirac $\cancel{CP}$ phase and the Jarlskog-Greenberg invariant \cite{jarlskog1, *greenberg1} predicted by the asymmetric texture are \cite{Rahat:2018sgs}:
\begin{align}
\delta_{CP} = 1.32 \pi, \qquad \mc J = -0.028,
\end{align}
compared to the latest PDG global fit
$\delta_{CP}^{PDG} = 1.37 \pm 0.17\pi$ at $1\sigma$ \cite{pdglive}. Hence, the sign of  low energy $CP$ violation and high energy baryon asymmetry would be consistent with data if the generated asymmetry is positive for negative $\delta$. 

In our analysis, all four regions of case 1 yield positive baryon asymmetry for negative $\delta$, whereas for case 2, it happens for $0.1 \leq a \leq 0.35$ and $a \geq 3.5$. The remaining regions of case 2 results in positive asymmetry for positive $\delta$.

With $\delta \simeq -78^\circ$, the sign of the Majorana invariants \cite{Bilenky:1980cx, *Schechter:1980gr, *Doi:1980yb, *schechter1982neutrinoless} is also fixed and the invariants are given by \cite{Perez:2020nqq}
\begin{align}
    \mc I_1 = -0.106, \quad \mc I_2 = -0.011.
\end{align}
Although there are still no strict bound on the Majorana phases from current experiments \cite{Ge:2016tfx, *Minakata:2014jba}, the prediction for $\delta_{CP}$ in the asymmetric texture is consistent with the current PDG fit. Recently $\delta_{CP} = 0$ has been excluded by the T2K experiment at $3\sigma$ level \cite{Abe:2019vii}, and upcoming experiments DUNE \cite{Abi:2020wmh} and Hyper-K \cite{Abe:2018uyc} are expected to measure $\delta_{CP}$ with $5\sigma$ precision in the next decade.
\section{Concluding Remarks} \label{sec:6}
In this paper we have investigated non-resonant thermal leptogenesis in the context of the asymmetric texture in the $SU(5) \times \mc T_{13} \times \mc Z_{12}$ model proposed in Refs.~\cite{Rahat:2018sgs,Perez:2019aqq,Perez:2020nqq}. Baryon asymmetry is generated through the decay of four right-handed Majorana neutrinos and is intimately related to the single $\cancel{CP}$ phase in the TBM seesaw mixing. The sign and magnitude of the asymmetry constrains the parameter space of the model and resolves the sign ambiguity in the TBM phase. 

We have shown that low energy $CP$ violation does not yield any high energy $CP$ asymmetry when all charged-lepton flavors are considered equivalent. This happens because the only source of $CP$ violation, the TBM phase, is introduced in a diagonal matrix and does not enter in the calculation of the $CP$ asymmetry. However, flavored leptogenesis remains viable and the low energy $\cancel{CP}$ phase generates non-vanishing $CP$ asymmetry.

The conventional analysis of thermal leptogenesis assumes a hierarchical mass spectrum of the Majorana neutrinos, where the asymmetry generated by the heavier ones are washed out completely and only the decay of the lightest one yields the baryon asymmetry. Such a generic picture  does not apply to the model discussed in this paper as the mass spectrum is non-hierarchical in the parameter space of interest. We have considered the decay of all four Majorana neutrinos in flavored leptogenesis and the resulting density matrix equations have been solved numerically. Our calculation of the baryon asymmetry relates the previously undetermined parameters of the model and determines the masses of the Majorana neutrinos to be  of $\mc O(10^{11} - 10^{12})$ $\text{GeV}$. 

We have illustrated that the unresolved sign of the TBM phase is related to the sign of the baryon asymmetry. Requiring the baryon asymmetry to be positive determines the sign of the TBM phase. Of the two variants of the vacuum expectation values considered in this paper, one case fixes the sign of the TBM phase to be consistent with the current experimental data in the whole parameter space away from mass degeneracies, whereas in the  other case both signs generate positive asymmetry in different parts of the parameter space.


The discussion in this paper has been limited to thermal leptogenesis in the strong washout regime, where the dynamics are described by density matrix equations. However, the mass spectrum in the model offers richer phenomenology. The parameter space includes regions of nearly degenerate Majorana neutrinos, where the $CP$ asymmetry is enhanced resonantly. This can further lower the required mass scale for reproducing the observed baryon asymmetry, even to the $\text{TeV}$ scale. The discussion of resonant leptogenesis in this model remains out of the scope of this paper and will be addressed in a future work. 
\begin{acknowledgements}
M.H.R. thanks Dr. Chee Sheng Fong, Dr. M. Jay P\'erez, Dr. Pierre Ramond, Dr. Alexander Stuart, Bin Xu and Dr. Jue Zhang for helpful discussion at different stages of the work and comments on the manuscript. He also thanks the referee for their valuable comments and suggestions. This research was partially supported by the U.S. Department of Energy under grant number DE-SC0010296.
\end{acknowledgements}





\newpage
\appendix

\section{The case for the $SU(5) \times \mc T_{13} \times \mc Z_{14}$ Model} \label{app:Z14}
In this appendix we show that the $SU(5) \times \mc T_{13} \times \mc Z_{14}$ model discussed in Ref.~\cite{Perez:2020nqq} does not yield successful leptogenesis even for the three flavor approximation. The particle content and their transformation properties are listed in Table \ref{table:modelZ14}: 
\begin{table}[ht]\centering
\renewcommand\arraystretch{1.1}
\begin{tabularx}{\textwidth}{@{}l | Y Y Y Y Y Y Y Y @{}}
\toprule
     & $F$ & $\bar{N}$ & $\bar{N}_4$ & $\bar{H}_{\g{5}}$ & $\Lambda$  & $\varphi_{\mc A}$ & $\varphi_{\mc B}$ & $\varphi_z$ \\ 
\hline
$SU(5)$    & $\overline{ \g{5}}$ & $\g{1}$ & ${ \g{1}}$ & ${\g{5}}$ & $\g{1}$ & ${ \g{1}}$ & $\g{1}$ & $\g{1}$ \\ 
$\mc T_{13}$ &  $\g{3}_1$ & $\g{3}_2$ & $\g{1}$ & $\g{1}$ & $\gb{3}_1$  & $\gb{3}_2$ & $\g{3}_2$ & $\gb{3}_2$  \\ 
$\mathcal{Z}_{14}$    & $\g{\eta}$ & $\g{\eta^5}$ & $\g{\eta^7}$ & $\g{\eta^{11}}$ & $\g{\eta^{2}}$  & $\g{\eta^{11}}$ & $\g{\eta^4}$ & $\g{\eta^2}$   \\ 
\botrule
\end{tabularx} 
\caption{Charge assignments of matter, Higgs, messenger and familon fields in the seesaw sector. Here $\eta^{14} = 1$. The $\mc Z_{14}$ `shaping' symmetry is required to prevent unwanted tree-level operators.}
\label{table:modelZ14}
\end{table}

The familon $\varphi_v$ of the $SU(5) \times \mc T_{13} \times \mc Z_{12}$ model is replaced by the familon $\varphi_z$, which contributes a term $\bar{N} \bar{N}_4\varphi_z$, replacing the term $\bar{N} \overline{\Lambda}\varphi_{v}$ in the Lagrangian Eq.~\eqref{seesawlag}. In Ref.~\cite{Perez:2020nqq}, the VEV of $\varphi_z$ was determined to be $\vev{\varphi_{z}} \equiv m_{bz}^2 (b_1^{-1}, -2b_3^{-1}, b_2^{-1})$, where the parameter $m_{bz}$ is related to the other parameters by
\begin{align*}
    \frac{6m_{bz}^4 + mb_1 b_2 b_3}{m b_1 b_2 b_3} = \frac{1}{0.48} \equiv k
\end{align*}
using oscillation data.

For simplicity we set $b_1 = b_2 = b_3 = b$. This results in a simpler Dirac Yukawa matrix
\begin{align}
    Y^{(0)} &\equiv \frac{\sqrt{b m_\nu}}{\vev{\bar{H}_\g{5}}} \left(
\begin{array}{cccc}
 0 & 1 & 0 & 0  \\[0.5em]
 1 & 0 & 0 & 0 \\[0.5em]
 0 & 0 & -e^{i \delta} & 0 \\
\end{array}
\right). \label{Y0mat14}
\end{align}
The Majorana mass matrix gets contribution from the new term $\bar{N} \bar{N}_4\varphi_z$ and can be written as
\begin{align}
    \mc M &\equiv  b \left(
\begin{array}{cccc}
 0 & 1 & 1 & t  \\
 1 & 0 & 1 & -2t \\
 1 & 1 & 0 & t \\
 t & -2t & t & a
\end{array}
\right),
\end{align}
where $a \equiv \frac{m}{b}$ and $t \equiv \sqrt{\frac{a(k-1)}{6}}$. Its Takagi factorization yields
$\mc M = \mc U_m\ \mc D_m\ \mc U_m$, where
\begin{align}
    \mc D_m = \text{diag}(M_1, M_2, M_3, M_4),
\end{align}
\begin{align}
    \mc U_m &\equiv   \left(
\begin{array}{cccc}
 -\frac{i}{\sqrt{2}} &  \frac{1}{\sqrt{3}} & \frac{i\ t}{\sqrt{(2k-M_3-1)a - (M_3-1)}} & \frac{ t}{\sqrt{(2k+M_4-1)a + (M_4+1)}}  \\
 0 & \frac{1}{\sqrt{3}} & \frac{-2i\ t}{\sqrt{(2k-M_3-1)a - (M_3-1)}} & \frac{-2 t}{\sqrt{(2k+M_4-1)a + (M_4+1)}} \\
 \frac{i}{\sqrt{2}} & \frac{1}{\sqrt{3}} & \frac{i\ t}{\sqrt{(2k-M_3-1)a - (M_3-1)}} & \frac{ t}{\sqrt{(2k+M_4-1)a + (M_4+1)}} \\
 0 & 0 & \frac{-i\ (M_3-1)}{\sqrt{(2k-M_3-1)a - (M_3-1)}} & \frac{ M_4 + 1}{\sqrt{(2k+M_4-1)a + (M_4+1)}}
\end{array}
\right),
\end{align}
and 
\begin{align}
    M_1 = b,\ M_3 &= \frac{b}{2} \left(\sqrt{(a-1)^2+4ak}-(a-1)\right), \nonumber \\
    \ M_2 = 2b, \ M_4 &= \frac{b}{2} \left(\sqrt{(a-1)^2+4ak}+(a-1)\right).
\end{align}

The Neutrino Yukawa matrix in the weak basis is defined as
\begin{align}
    Y_\nu = \mc U^{(-1)^\dagger} Y^{(0)} \mc U_m^*. \label{Ynu}
\end{align}
We now show that the $CP$ asymmetry vanishes in this model. 
\vspace{0.3cm}

The $CP$ asymmetry in the three flavor approximation is calculated from the diagonal elements of Eq.~\eqref{epsidm} and depends on the terms $\text{Im}[(Y_\nu^*)_{\alpha i} (Y_\nu)_{\alpha j} (Y_\nu^\dagger Y_\nu)_{ij}]$ and $\text{Im}[(Y_\nu^*)_{\alpha i} (Y_\nu)_{\alpha j} (Y_\nu^\dagger Y_\nu)_{ji}]$, where $j \neq i$. 
\vspace{0.2cm}

Explicitly calculating $Y_\nu^\dagger Y_\nu$, we see that the nonzero off-diagonal elements are $(34)$ and $(43)$ and they are imaginary. Hence the $CP$ asymmetry is zero unless either $i=3, j=4$ or $i=4, j=3$. 
\vspace{0.3cm}

Consider the case $i=3, j=4$. Since $(Y_\nu^\dagger Y_\nu)_{34}$ is imaginary, the $CP$ asymmetry would vanish if $(Y_\nu^*)_{\alpha 3} (Y_\nu)_{\alpha 4}$ is imaginary.
\vspace{0.2cm}

The nonzero elements of $Y^{(0)}$ are in the $(12)$, $(21)$ and $(33)$ position, where the first two are real and the last one is complex. From Eq.~\eqref{Ynu}, we can write
\begin{align} \label{proveimag}
    (Y_\nu^*)_{\alpha 3} &= (\mc U^{(-1)^T})_{\alpha k} Y^{(0)^*}_{k l} (\mc U_m)_{l3}, \\
    (Y_\nu)_{\alpha 4} &= (\mc U^{(-1)^\dagger})_{\alpha m} Y^{(0)}_{m n} (\mc U_m^*)_{n4}, 
\end{align}
where $\{k,l\}$ and $\{m,n\}$ can be $\{1,2\}$ or $\{2,1\}$ or $\{3,3\}$. Thus, the product 
\begin{align}
(Y_\nu^*)_{\alpha 3} (Y_\nu)_{\alpha 4} = \sum_{k,l,m,n} (\mc U^{(-1)^T})_{\alpha k} Y^{(0)^*}_{k l} (\mc U_m)_{l3} (\mc U^{(-1)^\dagger})_{\alpha m} Y^{(0)}_{m n} (\mc U_m^*)_{n4}
\end{align}
is the sum of nine terms:
\begin{align*}
    \text{term 1:}\quad \{k,l\} = \{1,2\}, \{m,n\} = \{1,2\}, \\
    \text{term 2:}\quad \{k,l\} = \{1,2\}, \{m,n\} = \{2,1\}, \\
    \text{term 3:}\quad \{k,l\} = \{2,1\}, \{m,n\} = \{1,2\}, \\
    \text{term 4:}\quad \{k,l\} = \{2,1\}, \{m,n\} = \{2,1\}, \\
    \text{term 5:}\quad \{k,l\} = \{3,3\}, \{m,n\} = \{3,3\}, \\
    \text{term 6:}\quad \{k,l\} = \{1,2\}, \{m,n\} = \{3,3\}, \\
    \text{term 7:}\quad \{k,l\} = \{3,3\}, \{m,n\} = \{1,2\}, \\
    \text{term 8:}\quad \{k,l\} = \{2,1\}, \{m,n\} = \{3,3\}, \\
    \text{term 9:}\quad \{k,l\} = \{3,3\}, \{m,n\} = \{2,1\}. 
\end{align*}

In Eq.~\eqref{proveimag}, the elements of $\mc U^{(-1)}$ are always real. Notice that $(\mc U_m)_{l3}$ is imaginary and $(\mc U_m^*)_{n4}$ is real for any $l$ and $n$. Moreover, $(\mc U_m)_{13} = -\frac{1}{2} (\mc U_m)_{23} = (\mc U_m)_{33}$ and $(\mc U_m)_{14} = -\frac{1}{2} (\mc U_m)_{24} = (\mc U_m)_{34}$. 
\vspace{0.2cm}

In terms $1-4$, the only imaginary component is $(\mc U_m)_{l3}$ and all other components are real. Hence the terms $1-4$ are imaginary. 
\vspace{0.2cm}

For term $5$, $Y^{(0)}_{33}$ contributes $e^{i\delta}$ and $Y^{(0)*}_{33}$ contributes $e^{-i \delta}$, thus making the product $Y^{(0)}_{33}Y^{(0)*}_{33}$ real. $\mc (U_m)_{33}$ is imaginary and all other components are real. Therefore the term $5$ is imaginary. 
\vspace{0.2cm}

Next, we consider the terms $6-7$. Explicitly writing their sum, we get
\begin{align*}
    &(\mc U^{(-1)^T})_{\alpha 1} Y^{(0)^*}_{1 2} (\mc U_m)_{23} (\mc U^{(-1)^\dagger})_{\alpha 3} Y^{(0)}_{3 3} (\mc U_m^*)_{34} + (\mc U^{(-1)^T})_{\alpha 3} Y^{(0)^*}_{3 3} (\mc U_m)_{33} (\mc U^{(-1)^\dagger})_{\alpha 1} Y^{(0)}_{1 2} (\mc U_m^*)_{24} \\
    = &-\frac{1}{2}(\mc U^{(-1)^T})_{\alpha 1} (\mc U^{(-1)^T})_{\alpha 3} \left|Y^{(0)}_{33}\right| (\mc U_m)_{23} (\mc U_m^*)_{24} (e^{i\delta} + e^{-i\delta}) \\
    = &- (\mc U^{(-1)^T})_{\alpha 1} (\mc U^{(-1)^T})_{\alpha 3} |Y^{(0)}_{33}| (\mc U_m)_{23} (\mc U_m^*)_{24} \cos{\delta}.
\end{align*}
In the second line, we have used $-\frac{1}{2} (\mc U_m)_{23} = (\mc U_m)_{33}$ and $-\frac{1}{2} (\mc U_m)_{24} = (\mc U_m)_{34}$ and the fact that $\mc U^{(-1)}$ is real. The third line is imaginary since $(\mc U_m)_{23}$ is imaginary and all other components are real. Hence the sum of the terms $6-7$ is imaginary. 
\vspace{0.2cm}

Similar arguments can be used to show that the sum of the terms $8-9$ are imaginary. Therefore, the right hand side of Eq.~\eqref{proveimag} is imaginary and $\text{Im}[(Y_\nu^*)_{\alpha i} (Y_\nu)_{\alpha j} (Y_\nu^\dagger Y_\nu)_{ij}]$ is zero.
\vspace{0.3cm}

The case for $i = 4, j = 3$ is the complex conjugate of the case $i=3, j=4$ and thus implies the same conclusion. 

\vspace{0.3cm}
Hence the $SU(5) \times \mc T_{13} \times \mc Z_{14}$ model does not yield successful leptogenesis for the simple choice of VEV $b_1 = b_2 = b_3 = b$ in the three flavor approximation. More general vacuum expectation values, for example the two cases discussed in Section 5, can result in non-vanishing $CP$ asymmetry. It is beyond the scope of the present paper and will be pursued in a future work.
\newpage

\section{The case for $f=1$ in the $SU(5) \times \mc T_{13} \times \mc Z_{12}$ Model} \label{app:feq1}
In the $SU(5) \times \mc T_{13} \times \mc Z_{12}$ model, setting $f=1$, i.e., $b_1 = b_2 = b_3 \equiv b$ implies that two of the mass eigenvalues are same:
\begin{align}
    M_1 = b, \quad M_2 = b, \quad M_3 = 2b, \quad M_4 = ab.
\end{align}
In the context of leptogenesis, when two right handed neutrinos have the same mass, their interference with each other yields zero $CP$ asymmetry. However, their interaction with the other right handed neutrinos can, in general, generate nonzero $CP$ asymmetry and may result in successful leptogenesis. In this Appendix we explore this possibility.

In this case the Dirac Yukawa matrix $Y^{(0)}$, the Majorana matrix $\mc M$ and the unitary matrix $\mc U_m$ can be read either from Eqs.~ \eqref{Y0}, \eqref{Majo} and \eqref{Unitm}, or from Eqs.~ \eqref{Y0new}, \eqref{Majonew} and \eqref{Unitmnew}, by setting $f=1$. For $\delta=\mp 78^\circ$, this yields  
\begin{align}
    Y_\nu = \frac{\sqrt{b m_\nu}}{v}\left(
\begin{array}{cccc}
  -0.7441 i & \mp 0.1426+0.3414 i & 0.5471\, \mp 0.1009 i & (0.5890\, \pm 0.0515 i) \sqrt{a} \\
  -0.6604 i & \pm 0.0404\, -0.4240 i & -0.6178 \pm 0.0285 i & (0.2601\, \mp 0.0146 i) \sqrt{a} \\
  -0.1013 i & \pm 0.7848\, +0.2567 i & 0.0091\, \pm 0.5549 i & (0.1561\, \mp 0.2835 i) \sqrt{a} \\
\end{array}
\right), 
\end{align}
and
\begin{align}
Y_\nu^\dagger Y_\nu = \frac{b m_\nu}{v^2} \left(
\begin{array}{cccc}
 1 & 0 & 0 & 0.6258 i \sqrt{a} \\
 0 & 1 & 0 & -0.3613 i \sqrt{a}  \\
 0 & 0 & 1 & 0 \\
 -0.6258 i \sqrt{a} & 0.3613 i \sqrt{a} & 0 & 0.5221 a \\
\end{array}
\right). \label{YYf1}
\end{align}
From explicit calculation, we see that the following relation holds:
\begin{align}
    \left[(Y_\nu)_{\alpha 1} (Y_\nu)^*_{\beta 4} + (Y_\nu)_{\alpha 4} (Y_\nu)^*_{\beta 1}\right] (Y_\nu^\dagger Y_\nu)_{4 1} = -\left[(Y_\nu)_{\alpha 2} (Y_\nu)^*_{\beta 4} + (Y_\nu)_{\alpha 4} (Y_\nu)^*_{\beta 2}\right] (Y_\nu^\dagger Y_\nu)_{4 2}. \label{relationf1}
\end{align}
Using this relation and the fact that $M_1 = M_2$ for $f=1$, we calculate the $CP$-asymmetries following Eq.~\eqref{epsidm}:
\begin{itemize}
    \item $\varepsilon^{(1)}_{\alpha \beta} = -\varepsilon^{(2)}_{\alpha \beta}$ \\
    For $\varepsilon^{(1)}_{\alpha \beta}$, the only nonzero off-diagonal elements in $Y^\dagger_\nu Y_\nu$ are $(Y^\dagger_\nu Y_\nu)_{14}$ and $(Y^\dagger_\nu Y_\nu)_{41}$. Hence in this case, $i=1$, $j=4$ is the only combination yielding nonzero terms. Since $(Y^\dagger_\nu Y_\nu)_{14} = -(Y^\dagger_\nu Y_\nu)_{41}$, Eq.~\eqref{epsidm} gives
    \begin{align}
        \varepsilon^{(1)}_{\alpha \beta} &= \frac{i \left[ (Y_\nu)_{\alpha 1} (Y_\nu^*)_{\beta 4}  + (Y_\nu^*)_{\beta 1} (Y_\nu)_{\alpha 4} \right] (Y^\dagger_\nu Y_\nu)_{41}}{16 \pi (Y^\dagger_\nu Y_\nu)_{11}} \ \left\{\zeta\left( \frac{x_4}{x_1} \right) -\xi\left( \frac{x_4}{x_1} \right)
    \right\}. \label{e1}
    \end{align}
    
    \noindent Similarly, for $\varepsilon^{(2)}_{\alpha \beta}$, only $i=2$, $j=4$ yields nonzero terms. Using $(Y^\dagger_\nu Y_\nu)_{24} = -(Y^\dagger_\nu Y_\nu)_{42}$, Eq.~\eqref{epsidm} gives
    \begin{align}
        \varepsilon^{(2)}_{\alpha \beta} &= \frac{i \left[ (Y_\nu)_{\alpha 2} (Y_\nu^*)_{\beta 4}  + (Y_\nu^*)_{\beta 2} (Y_\nu)_{\alpha 4} \right] (Y^\dagger_\nu Y_\nu)_{41}}{16 \pi (Y^\dagger_\nu Y_\nu)_{22}} \ \left\{\zeta\left( \frac{x_4}{x_2} \right) -\xi\left( \frac{x_4}{x_2} \right)
    \right\}. \label{e2}
    \end{align}
    Since $(Y_\nu^\dagger Y_\nu)_{11} = (Y_\nu^\dagger Y_\nu)_{22}$ and $x_1 = x_2$, using Eq.~\eqref{relationf1} in Eqs.~\eqref{e1} and \eqref{e2} results $\varepsilon^{(1)}_{\alpha \beta} = -\varepsilon^{(2)}_{\alpha \beta}$.  

    \item $\varepsilon^{(3)}_{\alpha \beta} = 0$\\
    Since $(Y_\nu^\dagger Y_\nu)_{3j} = 0$ for any $j \neq 3$, $\varepsilon^{(3)}_{\alpha \beta} = 0$.
    
    \item $\varepsilon^{(4)}_{\alpha \beta} = 0$\\
    In this case the nonzero terms correspond to $i=4$ and $j=1,2$. Using $(Y_\nu^\dagger Y_\nu)_{ij} = -(Y_\nu^\dagger Y_\nu)_{ji}$, Eq.~\eqref{epsidm} yields
    \begin{align}
        \varepsilon^{(4)}_{\alpha \beta} &= -\frac{i \left[ (Y_\nu)_{\alpha 4} (Y_\nu^*)_{\beta 1}  + (Y_\nu^*)_{\beta 4} (Y_\nu)_{\alpha 1} \right] (Y^\dagger_\nu Y_\nu)_{41}}{16 \pi (Y^\dagger_\nu Y_\nu)_{44}} \ \left\{\zeta\left( \frac{x_1}{x_4} \right) -\xi\left( \frac{x_1}{x_4} \right)
    \right\} \nonumber \\
    & -\frac{i \left[ (Y_\nu)_{\alpha 4} (Y_\nu^*)_{\beta 2}  + (Y_\nu^*)_{\beta 4} (Y_\nu)_{\alpha 2} \right] (Y^\dagger_\nu Y_\nu)_{42}}{16 \pi (Y^\dagger_\nu Y_\nu)_{44}} \ \left\{\zeta\left( \frac{x_2}{x_4} \right) -\xi\left( \frac{x_2}{x_4} \right)
    \right\}.
    \end{align}
Using $x_1 = x_2$ and Eq.~\eqref{relationf1}, this yields $\varepsilon^{(4)}_{\alpha \beta} = 0$.     
\end{itemize}

Since the masses of $N_1$ and $N_2$ are same, the number densities and their derivatives would be same: $N_{N_1} = N_{N_2}$ and $\frac{dN_{N_1}}{dz} = \frac{dN_{N_2}}{dz}$. Thus in Eq.~\eqref{eq:NBLdm}, the source term on the right hand side becomes
\begin{align*}
    -\sum_i \varepsilon^{(i)}_{\alpha \beta} D_i (N_{N_i} - N_{N_i}^{eq}) &= \left( \varepsilon^{(1)}_{\alpha \beta} \frac{dN_{N_1}}{dz} + \varepsilon^{(2)}_{\alpha \beta} \frac{dN_{N_2}}{dz} \right ) + \varepsilon^{(3)}_{\alpha \beta} \frac{dN_{N_3}}{dz} + \varepsilon^{(4)}_{\alpha \beta} \frac{dN_{N_4}}{dz} \\
    &= 0 + 0 + 0 = 0.
\end{align*}
The first two terms yield zero since $\varepsilon^{(1)}_{\alpha \beta} = -\varepsilon^{(2)}_{\alpha \beta}$ and $\frac{dN_{N_1}}{dz} = \frac{dN_{N_2}}{dz}$. The last two terms vanish since the $CP$ asymmetries are zero. 
\newpage

\section{Robustness of the results with respect to $\delta$} \label{app:robust}

The only source of low energy $CP$ violation in the asymmetric texture is the TBM phase $|\delta| = 78^\circ$, whose magnitude was determined in Ref.~\cite{Rahat:2018sgs} to match the reactor angle to its 2018 PDG central value \cite{tanabashi2018review}. In this appendix we investigate if this value of $\delta$ is contained in the range that reproduces all three PMNS angles within $3\sigma$ of their 2020 PDG central values and if the leptogenesis results derived in section \ref{sec:4} are robust with respect to the variation of $\delta$ within this range.

\vspace{0.5cm}

\begin{figure}[!ht]
    \centering
    \includegraphics[width=0.48\textwidth]{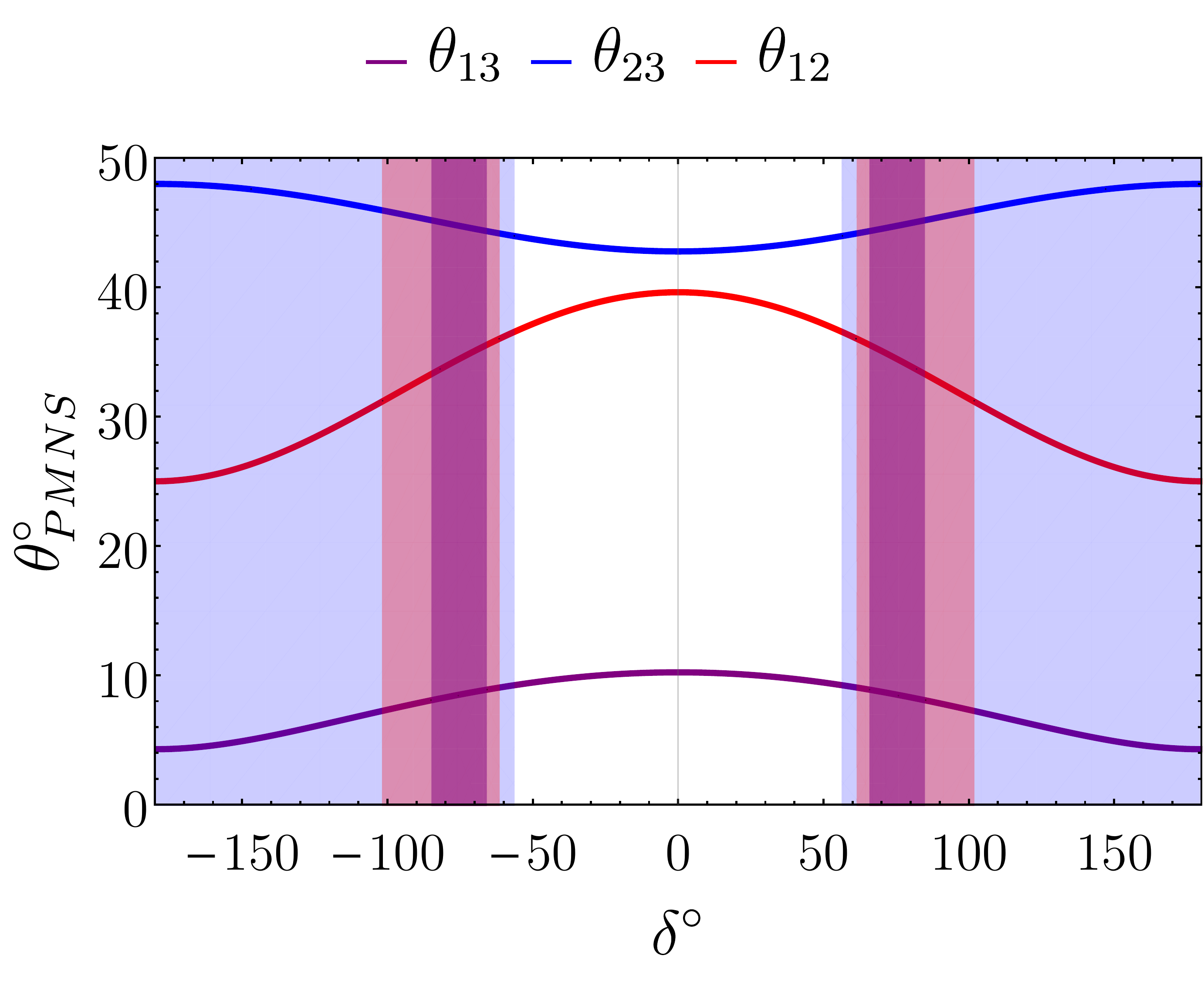}
\caption{Dependence of the PMNS angles on the TBM phase $\delta$. The shaded area represents the $3\sigma$ range of the angles from the 2020 PDG fit: $\sin^2{\theta_{13}} = 0.0218 \pm 0.0021$, $\sin^2{\theta_{23}} = 0.545 \pm 0.063$ and $\sin^2{\theta_{12}} = 0.307 \pm 0.039$ \cite{pdglive}. The common region where all three angles are within their $3\sigma$ fit is $66 ^\circ \leq \pm \delta \leq 85^\circ$. } 
\label{fig:pdg2020}
\end{figure}

The dependence of the PMNS angles on $\delta$ is shown in Figure \ref{fig:pdg2020}. The shaded regions represent the $3\sigma$ range of the latest PDG fit \cite{pdglive}.
For $66 ^\circ \leq \pm \delta \leq 85^\circ$, all three PMNS angles are within $3\sigma$ of their PDG central value. The corresponding range for the Dirac $\cancel{CP}$ phase is $1.27\pi \leq \mp \delta_{CP} \leq 1.35\pi$, consistent with the PDG fit $\delta_{CP}^{PDG} = 1.37 \pm 0.17 \pi$ \cite{pdglive}. Hence the results of Ref.~\cite{Rahat:2018sgs} are compatible with the 2020 PDG data.

\vspace{0.5cm}

In section \ref{sec:5}, we showed that the final $B-L$ asymmetry is proportional to $\sin \delta$. If $\delta$ is allowed to vary in $66^\circ \leq \delta \leq 85^\circ$, the Majorana masses required to reproduce the observed asymmetry will vary by a factor of $\mc O(\sin 66^\circ / \sin 78^\circ) \simeq \mc O(0.93)$ to $\mc O(\sin 85^\circ / \sin 78^\circ) \simeq \mc O(1.02)$. Hence the results are robust with respect to the variation in $\delta$.
\clearpage
\bibliography{leptogenesis}
\newpage
\bibliographystyle{apsrev4-1}

\end{document}